\documentclass[fleqn,twoside]{article}

\usepackage{amsmath}
\usepackage{amsfonts}
\usepackage{amssymb}
\usepackage{graphicx}%
\def\be{\begin{equation}}
\def\ee{\end{equation}}
\def\bi{\bibitem}

\begin{document}
\title{Euclidean wormholes with minimally coupled scalar fields}
\author{Soumendranath Ruz $^{1,*}$, Subhra Debnath$^{2,\dag}$, Abhik Kumar Sanyal$^{2,\ddag}$ and Bijan
Modak$^{1,\S}$}
\maketitle \noindent
\begin{center}
\noindent
$^{1}$ Dept. of Physics, University of Kalyani, West Bengal, India - 741235.\\
\noindent
$^{2}$ Dept. of Physics, Jangipur College, Murshidabad,
\noindent
West Bengal, India - 742213. \\

\end{center}
\footnotetext[1] {\noindent
Electronic address:\\
\noindent
$^{*}$ruzfromju@gmail.com\\
\noindent$^{\dag}$subhra\_ dbnth@yahoo.com\\
\noindent
$^{\ddag}$sanyal\_ ak@yahoo.com\\
\noindent
$^{\S}$bijanmodak@yahoo.co.in}
\begin{abstract}
A detailed study of quantum and semiclassical Euclidean wormholes for Einstein's theory with a minimally coupled scalar field has been performed for a class of potentials. Massless, constant, massive (quadratic in the scalar field) and inverse (linear) potentials admit Hawking and Page wormhole boundary condition both in the classically forbidden and allowed regions. Inverse quartic potential has been found to exhibit semiclassical wormhole configuration. Classical wormholes under suitable back-reaction leading to a finite radius of the throat, where strong energy condition is satisfied, have been found for the zero, constant, quadratic and exponential potentials. Treating such classical Euclidean wormholes as initial condition, late stage of cosmological evolution has been found to remain unaltered from standard Friedmann cosmology, except for the constant potential which under back-reaction produces a term like negative cosmological constant.

\end{abstract}
\noindent

\section{\bf{Introduction}}
Lot of research was initiated in wormhole physics only after Giddings and Strominger \cite{gs:1} for the first time presented an wormhole solution taking into account a third rank antisymmetric tensor field as the matter source and the rest of the world came to know that gravitational instantons could really exists. Wormholes, by then were considered as gravitational instantons, which are the saddle points of the Euclidean path integrals and as such are the solutions of the Euclidean field equations. Thereafter, research in wormhole physics was oriented (apart from traversable and non-traversable Lorentzian wormholes \cite{mt:2}) by and large in two directions. On one hand, people were motivated to find wormhole solutions for different types of matter fields \cite{l3} - \cite{l6} and on the other, physical consequences of wormhole dynamics were studied extensively \cite{h7} - \cite{h14}. In a nutshell, the outcome of all these works are the following. Firstly, it was learnt that quantum coherence is not really lost by the fact that wormholes connect two asymptotically flat or de-Sitter regions by a throat of radius of the order of Planck length. Next, microscopic wormholes might provide us with the mechanism that would solve the cosmological constant problem, while macroscopic wormholes might be responsible for the final stage of evaporation and complete disappearance of black hole. While these compelling results demand further investigation, the final outcome is disastrous, which is, not all type of matter fields admit wormhole solutions. If wormholes are considered seriously to be responsible for regularizing important physical parameters as stated above, then not only all type of matter fields but also pure gravity should admit wormhole solutions. This led Hawking and Page \cite{hp:p} to interpret wormholes in a different manner. They proposed that instead of considering wormholes as solutions to the classical field equations, it should be treated as solutions of the Wheeler-DeWitt (W-D) equation under the boundary conditions that the wave functional $\psi$ should be exponentially damped for large three geometry and it should be regular in some suitable way when the three geometry degenerates. The boundary condition mentioned above might not hold for the whole super-space, however, if it is satisfied in minisuperspace models, then wormholes are supposed to exist. Let us try to understand in brief how our usual conception on wormholes tallies with such boundary condition. W-D equation is independent of the lapse function and as such holds for both Euclidean and Lorentzian geometry. Solutions to the W-D equation are obtained both for classically allowed and forbidden regions depending on the signature of the potential. The solutions in the classically allowed regions are oscillatory and the corresponding states are not normalizable since the motion of the gravitational field is unbounded. For the Robertson-Walker (R-W) minisuperspace model, these solutions represent Friedmann solutions with unavoidable singularities. The solutions in the classically forbidden region are exponential in nature and represent Euclidean solutions. These solutions corresponding to the classically forbidden region might represent wormholes if the wave-functional, as already stated, is damped exponentially for large three geometry ensuring asymptotic flat regions which represents Euclidean space. Further, if it is regular as the three geometry degenerates, it assures a non-evolving throat of radius of the order of Planck length, instead of singularity. Hence it is clear that the proposal of Hawking and Page \cite{hp:p} is perfectly in tune with the preoccupied conception of wormholes.
\par
Along with their proposal, Hawking and Page \cite{hp:p} also presented a couple of such wormhole solutions corresponding to massless and massive scalar fields. For the massless case, under suitable transformation of variables, they \cite{hp:p} obtained a solution to the W-D equation as the product of two noninteracting harmonic oscillator wave functions with opposite energies that satisfies the boundary condition. The authors \cite{hp:p} however, did not indicate how much separation is required between the two harmonic oscillators to make such approximation to the W-D equation. For the massive case, their method of finding wormhole solution is so cumbersome that it really does not make much appeal. Further, they have not indicated how to find wormhole solutions for pure gravity. Later, Garay \cite{g:p} had explored the wormhole wave functional for a conformally coupled massless scalar field in the path integral method. Thereafter Coule \cite{c:c}, without going for the solutions of the W-D equation had shown that wormhole solutions might exist for a nonminimally coupled scalar field, just by studying the form of the potential. However, the main issue has not yet been settled, ie., whether all forms of matter fields really admit wormhole solution and a complete study in this regard does not exist in the literature. In this context, the motivation of the present paper is to take into account Einstein-Hilbert action with a minimally coupled scalar field and to explore quantum and semiclassical wormhole solutions for different form of the potentials, in the background of R-W minisuperspace model. It is important to mention that the radii of the throats of the wormholes which exist under semiclassical approximation are post-Planckian. It has been observed that wormhole boundary conditions \cite{hp:p} are satisfied in the semiclassical limit only for a limited form of potentials and for a limited class of operator ordering indices admitting back-reaction. Although it appears unrealistic, Kim \cite{kim} had suggested that if the wave functional well behaved for some operator ordering indices although divergence appears for different choice, it should still be considered to satisfy Hawking-Page boundary condition for wormholes.
\par
In the following section, we write the action for minimally coupled scalar theory of gravity, discuss our primary motivation in connection with some essential aspects of classical wormhole solutions which lead us to go for semiclassical approximation of the W-D equation. In section 3, we make WKB approximation to the W-D equation by expanding the phase of the wavefunctional in the power series of the gravitational constant $G$, or equivalently, ${m_{p}}^2$ where, $m_{p}$ is the reduced Planck's mass \cite{tp:i}, \cite{others}, instead of $\hbar$. In the process, one obtains Hamilton-Jacobi (H-J) equation for the source free gravity to the leading order of approximation. The H-J equation in turn reduces to the classical vacuum Einstein's equation under a suitable choice of time parameter. To the next order of approximation one obtains Tomonaga-Schwinger equation, which is essentially the functional Schr\"{o}dinger equation for the matter field propagation in the background of classical curved space. The underlying motivation is that, with this type of approximation the dynamics of the matter field is determined by the quantum field theory in curved space-time and when this equation is combined with vacuum Einstein's equation, a possible back-reaction might arise. If one chooses Hartle-Hawking type of wave-function \cite{give} for pure gravity, it is not difficult to see that the wavefunctional exponentially decays for large three geometry. Further, as the three geometry degenerates, the divergence due to the presence of Van-Vleck determinant can be regularized if back-reaction exists. In section 4, we take massless scalar field and a constant potential. In section 5, we take up power law potentials in the form of massive scalar field ($V_0\phi^2$), quartic potential ($\lambda\phi^4$) and inverse potentials ($\frac{V_0}{\phi^n}, n = 1, 3, 4$). In section 6, we take up exponential potential. In all the cases we first try to find quantum wormhole solutions by solving the W-D equation directly and then we explore semiclasssical wormhole solution following the technique mentioned above. Next, we present the classical wormhole equation and calculate the throat from the back-reaction in the cases for which semiclassical wavefunction admits back-reaction. Finally, we check if a viable classical cosmological evolution is admissible taking into account such back-reaction as initial condition.
\section{\bf{Action, motivation and the Wheeler-DeWitt equation}}

The gravitational action with a minimally coupled scalar field is,

\be\label{ac} S_c = \int d^4 x \sqrt{-g}\left[\frac{R}{16\pi G}-\frac{1}{2\pi^2}\left(\frac{1}{2}\phi_{,\alpha}\phi^{,\alpha}+V(\phi)\right)\right]-\frac{1}{8\pi G}\int_{\sum}d^3 x \sqrt{h}K .\ee

\noindent
Here $V(\phi)$ is an arbitrary potential and the surface term includes the determinant of the induced metric $h_{ij}$ along with the trace of the extrinsic curvature $K$. The trace energy tensor for the matter field under consideration is given by

\be T_{\mu\nu} = -\frac{2}{\sqrt{-g}}\frac{\delta S_{\phi}}{\delta g^{\mu\nu}} = \phi_{,\mu}\phi_{,\nu} - g_{\mu\nu}\Big[\frac{1}{2} \phi_{,\alpha}\phi^{,\alpha} + V(\phi)\Big],\ee
$S_{\phi}$ being the action for the matter field. The field equation for such an action may be computed as,

\be R_{\mu\nu} = 8\pi G [\phi_{,\mu}\phi_{,\nu} + g_{\mu\nu} V(\phi)]. \ee
It has been conjectured \cite{cg} that necessary (but not the sufficient) condition for a classical wormhole to exist is that the eigenvalues of the Ricci tensor must be negative somewhere on the manifold. This conjecture assures a real throat and so it is a necessary condition but since it does not assure asymptotic flat or de-Sitter space, therefore is not sufficient. Now, since $R_{\mu\nu}$ does not have negative eigenvalue for $V(\phi) > 0$, so clearly wormhole does not exist for real scalar in general in the Euclidean regime. For this reason Hawking and Page \cite{hp:p} conjectured wormhole as solution to W-D equation under appropriate boundary condition, mentioned in the introduction. Such boundary condition has been found to be satisfied for real massless and massive scalar fields \cite{hp:p}. We shall review these cases to show that negative eigenvalue of $R_{\mu\nu}$ exists under back-reaction following semiclassical approximation. In the following we elaborate this fact. In the Robertson-Walker minisuperspace model,

\be\label{RB} ds^2 = -dt^2+a(t)^2\left[\frac{dr^2}{1-kr^2}+r^2(d\theta^2+sin^2\theta\; d\phi^2)\right].\ee

\noindent The action (\ref{ac}) reduces to

\be S_c = M\int \left[-\frac{1}{2}a\dot a^2+\frac{k a}{2} +\frac{1}{M}\left(\frac{1}{2}\dot\phi^2-V(\phi)\right)a^3\right]dt .\ee

\noindent
Here $M = \frac{3\pi}{2G} = \frac{3\pi m_{p}^2}{2}$, $m_p$ being the reduced Planck mass and the curvature parameter $k = 0, \pm 1$, which stands for the flat, closed and open models respectively. The field equations are,

\be\label{1} \frac{\dot a^2}{a^2} + \frac{k}{a^2} = \frac{2}{M}\rho_{\phi}=\frac{2}{M}\left[ \frac{1}{2}\dot\phi^2 + V(\phi)\right] .\ee

\be\label{2} \ddot\phi + 3\frac{\dot a}{a}\dot\phi + V_{,\phi} = 0 .\ee

\noindent
Note that the other equation, viz., (1 1) = (2 2) = (3 3) is not relevant since it is not an independent one. Under Wick rotation $(t = i\tau)$ the field equations (\ref{1}) and (\ref{2}) take the form

\be\label{30} \frac{a_{,\tau}^2}{a^2} = \frac{k}{a^2} + \frac{2}{M}\left[\frac{1}{2}\phi_{,\tau}^2 -V(\phi)\right]\ee
\be\label{31} \phi_{,\tau\tau} + 3\frac{a_{,\tau}}{a}\phi_{,\tau} = V'(\phi).\ee

\noindent
Classical Euclidean wormholes, as mentioned, require two asymptotically flat spaces connected by a non-evolving throat where $a_{,\tau} = 0$ and is typically described by the form
\be\label{34} a_{,\tau}^2 = 1 - \frac{l^2}{a^n}, \;\;n > 0, \ee
where, $l$ and $n$ are constants. This form guarantees a real throat together with the condition $a_{,\tau}^2 > 0$ asymptotically, assuring asymptotic Euclidean (flat) space. For a better understanding let us take $n = 2$, to obtain a solution in the form
\be a^2 = l^2 + (\tau - \tau_0)^2.\ee
It is now clear that at $\tau = \tau_0$ a real throat $a_0 = l$ is found while as $\tau \rightarrow \pm \infty$, $a \rightarrow \infty$ and the wormhole boundary condition is satisfied. Cotsakis et al \cite{cot} had chosen an ansatz in the form $\phi_{,\tau} = \frac{l}{a^n}$ and found wormhole solution for real scalar field in the Euclidean domain which appeared due to some errors in sign being pointed out later by Coule \cite{coule}. As a result, (\ref{30}) and (\ref{31}) admit the above form provided the scalar field is imaginary in the Euclidean regime, ie., letting $\phi \rightarrow i\phi$, or $l \rightarrow il$ which are well known. Instead, a real scalar field (in the Lorentzian regime) being described by the following ansatz
\be\label{32} \dot \phi = \frac{l}{a^q},\ee
where, $q$ is a constant is more natural since similar form is  found as Noether conserved current $I = l = \dot\phi~a^3$ ($q = 3$) for cyclic scalar field $\phi$ (Note that under the assumption $\phi_{,\tau} = \frac{l}{a^q}$, one obtains $i\dot\phi~a^3 =$ conserved, which is unphysical). In view of the ansatz (\ref{32}), equation (\ref{2}) or (\ref{31}) may then be solved to obtain the form of the potential as,
\be\label{33} V = \frac{l^2(3-q)}{2qa^{2q}} \ee
which restricts the potential to be positive definite for $q \le 3$. In view of the above form of the potential (\ref{33}) equation (\ref{30}) may be expressed as,
\be\label{35} a_{,\tau}^2 = k - \frac{3 l^2}{M qa^{2(q-1)}}.\ee
It is apparent that equation (\ref{35}) is in the form of equation (\ref{34}) for $k = 1$ and so the necessary condition for wormhole to exist is satisfied. To understand the importance of this equation let us consider pure radiation instead of scalar field. Friedmann equation (\ref{1}) then reads

\be\label{fr1}\frac{\dot a^2}{a^2} + \frac{k}{a^2} = \frac{8\pi G}{3}\frac{\rho_{r0}}{a^4}\ee
$\rho_{r0}$ being a constant. Under Wick rotation $(t = i \tau)$, above equation reads
\be\label{fr2} a_{,\tau}^2 = 1 - \frac{8\pi G}{3}\frac{\rho_{r0}}{a^2}\ee
setting $k = +1$. The equation (\ref{fr2}) is clearly in the same form as (\ref{35}) under the choice $q = 2$. For pressureless dust again the Friedmann equation is
\be\label{fm1}\frac{\dot a^2}{a^2} + \frac{k}{a^2} = \frac{8\pi G}{3}\frac{\rho_{m0}}{a^3}\ee
$\rho_{m0}$ being a constant. Under Wick rotation $(t = i \tau)$, above equation reads
\be\label{fm2} a_{,\tau}^2 = 1 - \frac{8\pi G}{3}\frac{\rho_{m0}}{a}\ee
setting $k = +1$ and the equation is again in the same form as (\ref{35}) under the choice $q = 3/2$. Therefore both radiation and pressureless dust admit Euclidean wormhole boundary condition. Nevertheless, instead of radiation or pressureless dust, the early universe is vacuum dominated and so it is required to consider higher order curvature invariant terms or at least a scalar field. For a scalar field, which is our present consideration, the ansatz (\ref{32}) however under Wick rotation becomes $\phi_{,\tau} = i\frac{l}{a^q}$ and so the scalar field $\phi$ again becomes imaginary in the Euclidean domain. But then, we are not going to choose such an ansatz (\ref{32}), rather we show that a more general form of the ansatz (\ref{32}) is obtainable under semiclassical approximation as a back-reaction phenomena, if Hawking-Page wormhole boundary condition is satisfied. More clearly, one can calculate $<\rho_{\phi}>$ as a function of the scale factor $(a)$ under back-reaction appearing in semiclassical approximation of the W-D equation. In the process, for the real scalar field it is possible to obtain an equation having a more general form than equation (\ref{35}) without making a Wick rotation to the scalar field. Note that the only scalar field which exists in the form of a Higgs boson has recently been observed at LHC in the ATLAS \cite{atlas} and CMS \cite{cms} detectors applying $\sqrt{s} = 8 TeV$, which probes $10^{-18} cm$. Thus, the energy scale of the Higgs particle is much smaller than the scale of gravity and so it does not affect the geometry of the space-time. Hence, if a scalar is assumed to exist in the Planckian epoch, then we show that it might only leave its trace on the cosmological evolution through back-reaction. With this motivation, we proceed to make semi-classical approximation of the W-D equation, in the following section. The Hamilton constraint equation for the system under consideration is,
\be -\frac{1}{2M}\frac{P_a ^2}{a}+\frac{P_{\phi}^2}{2a^3}-\frac{M}{2}k a +a^3 V(\phi) = 0, \ee

\noindent
where, $ P_a $ and $ P_\phi $ are the corresponding momenta canonically conjugate to $a$ and $\phi$ and so the Wheeler-DeWitt (W-D) equation reads,

\be\label{WD} \left[\frac{\hbar^2}{2M}\left(\frac{\partial^2}{\partial a^2}+\frac{p}{a}\frac{\partial}{\partial a}\right)-\frac{M}{2}ka^2-\frac{\hbar^2}{2a^2}\frac{\partial^2}{\partial\phi^2} +a^4 V(\phi)\right]\mid\Psi> = 0,\ee

\noindent where, $p$ removes some of the operator ordering ambiguities. Let us remind that the W-D equation is independent of the lapse function and therefore its form is the same both in the Lorentzian and Euclidean space-time. In the following section we shall turn our attention to semiclassical approximation of the W-D equation (\ref{WD}).

\section{\bf{Semiclassical approximation.}}

For semiclassical approximation of the Wheeler-DeWitt equation (\ref{WD}), we choose
$\Psi(a,\phi) = \exp[\frac{i}{\hbar}S(a,\phi)]$ and substitute it along with its derivatives in the W-D equation (\ref{WD}) to get,

\be \Big[ \frac{i\hbar}{2M}S_{,aa} - \frac{1}{2M}S_{,a}^2 + \frac{i\hbar}{2M}p\frac{S_{,a}}{a} - \frac{M}{2}ka^2 - \frac{i\hbar}{2a^2}S_{,\phi\phi} + \frac{1}{2a^2}S_{,\phi}^2 + a^4V \Big]\psi = 0 .\ee

\noindent
Now let us expand the functional $S(a,\phi)$ in the power series of $M^{-1}$ (instead of $\hbar$) as, $S = MS_0 + S_1+M^{-1}S_2\cdots etc.$ and after substituting it (mentally) in the above equation (21), equate the coefficients of different orders of $M$ to zero. To the highest viz., $M^2$ order, we obtain,

\be \frac{\partial S_0}{\partial\phi} =0,\ee

\noindent which implies that $S_0$ is purely a functional of gravitational field ie., $S_0 = S_0(a)$. The next, ie., $M^1$ order term gives the following source free Einstein-Hamilton-Jacobi (EHJ) equation,

\be\label{S_0} \left(\frac{\partial S_0}{\partial a}\right)^2 + ka^2 =0. \ee

\noindent If one now identifies the derivative of the phase $S_0$ with classical momenta as $M\frac{\partial S_0}{\partial a} = P_a = -M a\dot a$, then EHJ equation (\ref{S_0}) reduces to the vacuum Einstein's equation, viz.,

\be\label{3} \dot a^2 + k = 0, \ee

\noindent or equivalently,

\be \frac{1}{2M}P_a^2 + \frac{M}{2}ka^2 = 0,\ee

\noindent under the following choice of time parameter,

\be\label{4} \frac{\partial}{\partial t} = - \frac{1}{a}\frac{\partial S_0}{\partial a}\frac{\partial}{\partial a}.\ee

\noindent At this stage we note that the curvature parameter $k = 0$ leads to static model and so we leave it. Next, since neither equation (\ref{S_0}) nor (\ref{3}) admits real solution of $S_0$ for $k = +1$, therefore apparently it is required to switch over to the Euclidean time following the transformation $t = i\tau$. In the process, equation (\ref{3}) reduces to $a_{\tau}^2 - k = 0$ and so real solution now exists for $k = +1$ in Euclidean section. If one now requires to obtain the time parameter (\ref{4}) from the action principle, then one has to rotate the EHJ function $S_0$ to the Euclidean plane following the transformation $S_0 = i I_{E_0}$, where, $I_{E_0}$ is the Euclidean Hamilton-Jacobi functional. Thus, equation (\ref{S_0}) reduces to

\be \left(\frac{\partial I_{E_0}}{\partial a}\right)^2 - ka^2 =0, \ee

\noindent and as mentioned, real solutions are now admissible for $k = +1$. Finally, if there exists a lower limit $a_0$ to $a$, then the analytic structure of Coleman-Hawking wormhole arises quite naturally. Thus we observe that the problem associated with the type of semiclassical approximation under consideration for a closed ($k = +1$) FRW model is resolved if one invokes wormhole configuration. It now remains to be shown - how this finite resolution limit of the order of Planck's length in the scale factor $a$ appears. The validity of the WKB approximation under consideration requires $\mid\frac{d\lambda_a}{d a}\mid << 1$, where, $\lambda_a$ is the de-Broglie wave-length for pure gravity. Identifying the derivative of the phase factor $S_0$ with the canonical momenta, the above statement reduces to $\mid\frac{d}{d a}(\frac{\hbar}{MS_{0,a}})\mid << 1$. In view of equation (\ref{S_0}) it implies $a >> \sqrt{\hbar/M}$ or $a >> 7.45\times 10^{-34}$ cm. The validity of WKB approximation under consideration, therefore requires any Planckian or post-Planckian value of the scale factor and so both the microscopic as well as macroscopic wormholes will possibly exist. Hence this method of semiclassical approximation is well posed to explain both the problems of vanishing of the cosmological constant and the final stage of evaporation and complete disappearance of black hole. However, these are not our present concern, rather we shall also find that open Friedmann model ($k = - 1$) does not satisfy Hawking-Page wormhole boundary condition. Now the next ($M^0$) order of approximation yields,

\be \frac{i\hbar}{2}\left[S_{0,aa} + \frac{p}{a}S_{0,a}\right] - S_{0,a}S_{1,a} - \frac{i\hbar}{2a^2}S_{1,\phi\phi} + \frac{1}{2a^2}S_{1,\phi}^2 + a^4V(\phi) = 0 \ee

\noindent
which, using equation (\ref{S_0}) and the time parameter defined in equation (\ref{4}) may be rearranged to obtain the following functional Schr\"{o}dinger equation, also known as Tomonaga-Schwinger equation, propagating in the background of curved space-time, viz.,

\be\label{7} -\frac{i\hbar}{a}\left(\frac{\partial S_0}{\partial a}\right)\frac{\partial f(a,\phi)}{\partial a}=i\hbar\frac{\partial f(a,\phi)}{\partial t}\Rightarrow \left[-\frac{\hbar^2}{2a^3}\frac{\partial^2}{\partial\phi^2}+a^3
V(\phi)\right]f(a,\phi) = \pm \hbar\sqrt k \frac{\partial f(a,\phi)}{\partial a },\ee

\noindent provided,

\be\label{5} \frac{\partial S_0}{\partial a}\left(\frac{\partial {\mathcal D}(a)}{\partial a}\right)-\frac{1}{2}\left(\frac{\partial^2 S_0}{\partial a^2}+\frac{p}{a}\frac{\partial S_0}{\partial a}\right) {\mathcal D}(a) = 0,\ee
\noindent where, $f(a, \phi)$ and ${\mathcal D}(a)$ are related by,

\be\label{41} f(a,\phi)={\mathcal D}(a)\exp{\left(\frac{i S_1}{\hbar}\right)}.\ee

\noindent Here, ${\mathcal D}(a)$ plays the role of Van-Vleck determinant and can be solved exactly in view equation (\ref{5}) to yield

\be\label{37} {\mathcal D}(a) = \mu^{-1} a^{\frac{p+1}{2}},\ee

\noindent where, $ \mu $ is the constant of integration. Upto this ($M^0$) order of approximation, the wave functional

\be \Psi(a,\phi) = \exp{\left[\frac{i}{\hbar}\left(MS_0+S_1\right)\right]}\ee takes the following form,

\be\label{6} \Psi(a,\phi) =  \mu a^{-\frac{p+1}{2}}\exp{\left[-\frac{M}{2\hbar}\sqrt k a^2\right]}f(a,\phi).\ee

\noindent To obtain the wormhole wave-functional we have chosen negative sign in the exponent of equation (\ref{6}), which is the Hartle-Hawking \cite{give} choice, as mentioned earlier. The exponent part is well behaved both for $a\rightarrow 0$ and $a\rightarrow \infty$. However, for $p+1 > 0$, the determinant diverges as $a\rightarrow 0$. If the solution of $f(a,\phi)$ can somehow control this divergence as $a\rightarrow 0$, then only one can expect wormhole configuration for other values of operator ordering index $p$. In the following sections, we shall attempt to find quantum wormholes as solutions to the W-D equation (\ref{WD}) and also try to find semiclassical wormholes in view of the wave-function (\ref{6}) by solving $f(a,\phi)$ explicitly in view of equation (\ref{7}) for a class of potentials $V(\phi)$. Throat of the wormhole is then found for potentials which admit back-reaction and finally, classical cosmological evolution will be studied considering back-reaction as an initial condition.

\section{Wormholes for zero and constant potentials}

\subsection{Case 1, $V(\phi) = 0$.}
As mentioned in the introduction, wormhole solution for vanishing potential was first found for axion coupled to gravity in Euclidean space-time \cite{gs:1}. These are classical charged wormholes. The same type of classical wormhole solutions were found later, where gravity is coupled to positive energy massless complex scalar field \cite{Burgess}. Hawking and Page \cite{hp:p} represented wormholes in a more general manner in quantum domain as the solution of Wheeler-DeWitt equation with appropriate boundary condition. They \cite{hp:p} also showed that wormhole solution exists for both massless and massive scalar field in this new approach. In the case of massless scalar field, solution of the W-D equation apparently does not admit Hawking-Page boundary condition in general. However, under change of variables $x = a\sinh\phi$ and $y = a\cosh\phi$, the W-D equation reduces to two harmonic oscillators with opposite signs of energy. These solutions are regular at the origin and damped at infinity for $p = 1$. Here, we first review the case and then show that wormhole boundary condition is obeyed under semiclassical approximation in a straight forward manner for arbitrary factor ordering index $p$.

\subsubsection{Quantum Wormhole}

For massless scalar field (vanishing potential) the W-D equation (\ref{WD}) may be reexpressed in the following form,

\be \left[\frac{\hbar^2 }{2M}\left(a^2 \frac{\partial^2}{\partial a^2} + pa\frac{\partial}{\partial a}\right)-\frac{M}{2}ka^4 \right]\Psi(a,\phi) = -\frac{\hbar^2}{2}\frac{\partial^2}{\partial\phi^2}\Psi(a,\phi).\ee
Under separation of variable $\Psi(a,\phi)= A(a)B(\phi)$, the above equation takes the form

\be \frac{1}{A}\left[\frac{\hbar^2 }{2M}\left(a^2 \frac{\partial^2A}{\partial a^2} + pa\frac{\partial A}{\partial a}\right)-\frac{M}{2}ka^4 A \right] = -\frac{\hbar^2}{2B}\frac{\partial^2B}{\partial\phi^2}= \omega^2,\ee

\noindent
where $\omega^2$ is the separation constant. Explicit solution of the wave function $\Psi(a, \phi)$ is therefore,
\be\label{36} \Psi(a,\phi) = (-1)^b\left( \frac{a^2M\sqrt k}{4\hbar} \right)^{\frac{1-p}{4}}\left[ C_1 I_{-c}\left(\frac{a^2M\sqrt k}{2\hbar} \right) \Gamma\left( 1-c \right) + C_2 (-1)^c \ I_c\left(\frac{a^2M\sqrt k}{2\hbar} \right) \Gamma\left( 1+c \right)  \right]B(\phi) \ee

\noindent
Here, $I_\alpha(x)$ is the modified Bessel function of first kind (see appendix for its properties). While, $C_1$ and $C_2$ are constants of integration and the constants $b$ and $c$ are the given by
\be b = \frac{1-p}{8} - \frac{c}{2} \hspace{.5 in} \text{and} \hspace{ .5 in} c = \frac{\sqrt{\hbar^2(p-1)^2 + 8M\omega^2 }}{4\hbar}\ee

\noindent
Since $B(\phi)$ is a regular oscillatory function, therefore it appears that the wave function (\ref{36}) does not satisfy wormhole boundary condition. Nonetheless, instead of assuming separation of variables, the W-D equation for massless scalar field with $p = k = 1$, under the change of variable $x = a \sin{h\phi}$ and $y = a \cos{h\phi}$ was expressed by Hawking and Page \cite{hp:p} in the following form (in the unit $\hbar = M = 1$),
\be\label{WDH} \left(\frac{\partial^2}{\partial y^2} - y^2 - \frac{\partial^2}{\partial x^2} + x^2\right)\psi(x, y) = 0,\ee

\noindent
The above equation represents two harmonic oscillators with opposite energy and so is well behaved at both ends, confirming the existence of wormhole for massless scalar field. This is possible because neither $A(a)$ nor $B(\phi)$ is regular but the combined solution $\Psi(a, \phi)$ turns out to be regular.

\subsubsection{Semiclassical Wormhole}

Having obtained quantum wormhole solution for vanishing potential $V(\phi) = 0$ for the operator ordering index $p = 1$, let us now proceed to find its fate under semiclassical approximation. Equation (\ref{7}) under separation of variables $f(a, \phi) = A(a) \Phi(\phi)$ takes the following form,

\be \pm 2\hbar\sqrt ka^3\frac{1}{A}\frac{\partial A}{\partial a} = -\hbar^2\frac{1}{\Phi}\frac{\partial^2\Phi}{\partial\phi^2} =  4C^2, \ee
where, $C^2$ is a separation constant. $A(a)$ and $\Phi(\phi)$ may now be solved immediately to obtain,

\be f(a,\phi)=\exp{\left(\mp\frac{C^2}{\hbar \sqrt k a^2}\right)}\Phi(\phi),\ee

\noindent where $\Phi$ is a regular oscillatory function. If we now restrict ourselves to the negative sign in the exponent of $f(a,\phi)$ (since for the other sign in $f(a, \phi)$ wormhole does not exists) then the wave functional $\Psi$ given in equation (\ref{6}) takes the following form,

\be\label{8} \Psi(a,\phi) =  \mu a^{-\frac{p+1}{2}}\exp{\left[-\frac{1}{\hbar}\left(\frac{M}{2}\sqrt k a^2+\frac{C^2}{\sqrt k a^2}\right)\right]}\Phi(\phi).\ee

\noindent Clearly the wave-functional is exponentially damped at large three geometry and regular at small three geometry in the case of closed $(k = +1)$ model for arbitrary factor ordering index $p$. The exponent of the matter wave-functional $f(a, \phi)$ not only controls the Van-Vleck determinant $\mathcal{D}(a)$ given in equation (\ref{37}) but also does the same to any arbitrary but finite oscillation of the scalar field $\phi$ in both ways $a \rightarrow 0$ and $a \rightarrow \infty$. Thus, massless scalar field admits Hawking-Page wormhole boundary condition in the semiclassical limit for arbitrary operator ordering index $p$ while quantum wormholes are realized only for $p = 1$. Note that for $k = -1$ the wave function becomes oscillatory leading to inhabitable singularity instead of wormhole.

\subsubsection{Back-reaction and the throat}

Let us now go a bit further. In order to show that the scalar field indeed induces a back-reaction on pure gravity and to calculate the throat, it is required to express the exponent of the solution (\ref{8}) as $\exp[\frac{i}{\hbar}S_t]$, fixing the curvature parameter to $k = +1$, since semiclassical wormhole for massless scalar field exists for closed model only. Therefore, we need to define the quantity $S_t$ as

\be\label{9} S_t = i\left[ \frac{M}{2} a^2  + \frac{C^2}{ a^2}\right],\ee

\noindent Now, taking derivative with respect to $a$ and upon squaring, equation (\ref{9}) reduces to

\be\label{38} S_{t,a}^2 = -M^2 a^2 + \frac{4C^2 M}{a^2} - \frac{4C^4}{a^6}.\ee

\noindent Now treating $S_{t,a} = - Ma\dot a$, as the classical momentum, the above equation (\ref{38}) may be viewed simply as the Einstein's equation along with the back-reaction term, viz.,

\be\label{10} \frac{\dot a^2}{a^2} + \frac{1}{a^2} = \frac{2}{M}<\rho_{\phi}> = \frac{2}{M}<\frac{1}{2}\dot\phi^2> = \frac{2}{M}\Big[\frac{2C^2}{a^6}  - \frac{2C^4}{M a^{10}}\Big] ,\ee
which under Wick rotation takes the form
\be a_{,\tau}^2 = 1- \frac{2}{M}\Big[\frac{2C^2}{a^4}  - \frac{2C^4}{M a^{8}}\Big] ,\ee

\noindent
which is in a much general form than  equation (\ref{34}) and clearly admits classical wormhole boundary condition. The throat of the wormhole can now be found setting $a_{,\tau} = 0$ and is given by
\be\label{22} a_0 = \left(\frac{2C^2}{M}\right)^{\frac{1}{4}}.\ee

\noindent Further, since $a_{,\tau}^2 > 0$, therefore asymptotically flat Euclidean space is assured. In the case under consideration $p_{\phi} = \rho_{\phi}$ and so an ultraviolet cutoff $a \ge (\frac{C^2}{M})^{1/4}$ is required to satisfy the strong energy conditions($\rho_{\phi} + 3 p_{\phi} \ge 0$). Size of the throat clearly shows the existence of an automatic ultraviolet cutoff and strong energy condition is obeyed. Depending on the value of the only parameter $C >> \sqrt{\hbar/2}$, the size of the throat may be anything of the order of post-Planckian length, which is the limit of semiclassical approximation as demonstrated in section 3. For example, if the constant $C$ is chosen of the order one then the throat is of the order of micrometer. It is also possible to obtain the so called classical wormhole boundary condition viz., equation (\ref{34}), if one sets the last term in equation (\ref{10}) to vanish which is only possible when the scale factor is sufficiently large ie., in the classical limit. In that case from the right hand side of equation (\ref{10}) one obtains $a^3 \dot\phi = 2C$, which is true at the classical level since $\phi$ is cyclic. Condition (\ref{32}) thus results in quite trivially.

\subsubsection{Late time Cosmic evolution}
Our analysis reveals that under semiclassical approximation, massless scalar field admits wormhole boundary condition for arbitrary factor ordering index (unlike quantum wormhole) corresponding to which a back-reaction exists leading to the gravitational potential
\be\label{U} U(a) = \frac{1}{a^2} - \frac{4C^2}{Ma^6} + \frac{4C^4}{M^2 a^{10}}.\ee

\noindent The throat of the wormhole assures that the scale factor $a$ is now bounded from below and so the gravitational potential $U(a)$ does not suffer from short distance instability, i.e., there exists an ultraviolet cutoff. Likewise, it is also clearly evident from the above equation (\ref{U}) that $U(a)$ is free from long distance instability. The asymptotic de-Sitter/flat space on the other hand assures an early inflationary epoch. Inflation must have ended by $10^{-32 \pm 6}s$ to give way to hot big bang before $T \sim 100 GeV$, for free quarks to exist restoring electroweak phase transition and to validate standard Nucleosynthesis. Thus inflation is purely a quantum phenomena and so classical field equations might not always exhibit such behaviour. It is now important to see how such initial wormhole boundary condition which sets up a throat affects cosmological evolution. For the purpose, it is required to add some contributions from radiation immediately after the asymptotic flat/de Sitter space ($k = 0$) has been arrived at and also some contribution from matter at the late stage. Hence, the classical field equations, which we need to solve are

\be \frac{\dot a^2}{a^2} = \frac{4}{M}\left[\frac{C^2}{a^6} - \frac{C^4}{M a^{10}} + \pi^2 \left(\frac{\rho_{ro}}{a^4} + \frac{\rho_{mo}}{a^3}\right)\right]\ee
\be a^3\dot \phi = 2C\ee
where wormhole boundary condition has been incorporated. In the above $\rho_{ro}$ and $\rho_{mo}$ are the amount of radiation and the matter available at the present epoch. Clearly as Universe expands, the contributions from the first and second terms are negligible and the Universe evolves like the usual Friedmann model with $a \propto t^{1/2}$ in the radiation era and $a \propto t^{2/3}$ in the matter dominated era. As a result, Baryogenesis, Nucleosynthesis, LSS (large scale structure) along with WMAP redshift data on matter-radiation equality and decoupling remain unaltered. However, late time cosmic acceleration definitely requires some form of dark energy, which has not been considered here. Thus massless scalar field turns out to be a good candidate to explain the cosmological evolution.

\subsection{$V(\phi) = V_0$, where $V_0$ is a constant.}

Lee \cite{Lee} had shown that the quantum theory of a complex scalar field with constant potential admits wormhole boundary condition provided $V_0$ has a minima, indicating that the size of the wormhole must be less than horizon length. Here we study the case for a real scalar field.

\subsubsection{Quantum wormhole}

W-D equation (\ref{WD}) may now be expressed in the form,

\be\label{wd2} \left[\frac{\hbar^2 }{2M}\left(a^2 \frac{\partial^2}{\partial a^2} + pa\frac{\partial}{\partial a}\right)-\frac{M}{2}ka^4 + a^6 V_0\right]\Psi(a,\phi) = -\frac{\hbar^2}{2}\frac{\partial^2}{\partial\phi^2}\Psi(a,\phi).\ee
Under separation of variables $\Psi(a,\phi)= A(a)B(\phi)$, the above equation reads

\be \frac{1}{A}\left[\frac{\hbar^2 }{2M}\left(a^2 \frac{\partial^2A}{\partial a^2} + pa\frac{\partial A}{\partial a}\right)-\frac{M}{2}ka^4 A + a^6 V_0 A \right] = -\frac{\hbar^2}{2B}\frac{\partial^2B}{\partial\phi^2}=\omega^2,\ee
$\omega^2$ being the separation constant. The gravitational part of the equation is solvable only for $\omega^2=0$ and for some particular value of $ p $. For $ p = -1 $ or $ p = 3 $ we can write solutions in the following form,
\be \Psi = {Ai}\left(\frac{2^{\frac{2}{3}}\left(\frac{kM^2}{4\hbar^2} - \frac{a^2MV_0}{2\hbar^2}\right)}{\left(\frac{-M V_0}{\hbar^2}\right)^{\frac{2}{3}}}\right)C_3B(\phi) + {Bi}\left(\frac{2^{\frac{2}{3}}\left(\frac{kM^2}{4\hbar^2} - \frac{a^2MV_0}{2\hbar^2}\right)}{\left(\frac{-M V_0}{\hbar^2}\right)^{\frac{2}{3}}}\right)C_4 B(\phi)\ee

and
\be \Psi = \frac{1}{a^2}{Ai}\left(\frac{2^{\frac{2}{3}}\left(\frac{kM^2}{4\hbar^2} - \frac{a^2MV_0}{2\hbar^2}\right)}{\left(\frac{-M V_0}{\hbar^2}\right)^{\frac{2}{3}}}\right)C_5B(\phi) + \frac{1}{a^2}{Bi}\left(\frac{2^{\frac{2}{3}}\left(\frac{kM^2}{4\hbar^2} - \frac{a^2MV_0}{2\hbar^2}\right)}{\left(\frac{-M V_0}{\hbar^2}\right)^{\frac{2}{3}}}\right)C_6 B(\phi)\ee

\noindent
where ${Ai(x)}$ and ${Bi(x)}$ are Airy functions of the first and second kind respectively, while, $ C_3,C_4,C_5,C_6 $ are constants. Airy functions with negative argument are highly oscillatory (see appendix) leading to Lorentzian regime with unavoidable singularity and wormhole boundary condition therefore is not satisfied. Nevertheless, instead of attempting solution under the assumption of separation of variables, the Wheeler-deWitt equation (\ref{wd2}) under the transformation $x = a\sin h\phi$, $y = a \cos h\phi$ and for $p = k = 1$ reduces to

\be \left[\frac{\partial^2}{\partial y^2} -y^2 - \frac{\partial^2}{\partial x^2} + x^2 + 2(y^2 - x^2)^2 V_0\right]\Psi = 0,\ee
which represents coupled harmonic oscillator with opposite energies and is well behaved at both ends. Thus, here again wormhole boundary condition is satisfied for $p = 1$.

\subsubsection{Semiclassical Wormhole}

Equation (\ref{7}) can again be solved for $f(a,\phi)$ using the method of separation of variables and for this particular case it can be written in the following form,
\be \pm \frac{1}{A(a)}\frac{\partial A(a)}{\partial a} = \frac{2C^2}{\hbar\sqrt k}\frac{1}{a^3} + \frac{V_0}{\hbar\sqrt k}a^3 \hspace{.3 in} \text{and} \hspace{.3 in}  -\hbar^2\frac{1}{\Phi(\phi)}\frac{\partial^2\Phi(\phi)}{\partial\phi^2} = 4C^2. \ee

\noindent
We have chosen the same separation constant $4C^2$ as in the previous case to make a comparison of the throat of the wormhole. Equations (56) may be solved and the explicit form of the wave-functional in view of equation (\ref{6}) is expressed as

\be\label{WF} \Psi(a,\phi) = \mu a^{-\frac{p+1}{2}}\exp{\left[-\frac{1}{\hbar}\left(\frac{M}{2}\sqrt k a^2 + \frac{C^2}{\sqrt k a^2} + \frac{V_0 a^4}{4\sqrt k}\right)\right]}\Phi(\phi),\ee

\noindent where, $\Phi(\phi)$ admits same type of oscillatory solution as before. The above form of the wavefunction is again well behaved i.e., regular at $a \rightarrow 0$ and damped out exponentially as $a \rightarrow \infty$ for $k = +1$ and for all values of the operator ordering index $p$. The exponent here again can control both the Van-Vleck determinant and arbitrary oscillations appearing in $\Phi(\phi)$. Thus $\Psi(a, \phi)$ obtained in (\ref{WF}) admits wormhole boundary condition.

\subsubsection{Back-reaction and the throat}

Proceeding as in the earlier case, ie., expressing the exponent of the solution (\ref{WF}) as $\exp[\frac{i}{\hbar}S_t]$, we obtain
\be\label{br} S_{t,a}^2 = -M^2 a^2  - \frac{4C^4}{a^6} + \frac{4 M C^2}{a^2} + 4V_0 C^2 - 2M V_0 a^4,\ee
neglecting higher order term in the scale factor (viz.,$~ a^6$) since such term does not contribute to the throat. Hence, Einstein's equation with back-reaction now takes the form,
\be\label{13} \frac{\dot a^2}{a^2}+\frac{1}{a^2}= \frac{2}{M}<\rho_{\phi}> = \frac{2}{M}\left[ \frac{2C^2}{a^6} -\frac{2C^4}{M a^{10}} + \frac{2V_0 C^2}{M a^4} - V_0  \right], \ee
which under Wick rotation reads
\be\label{Wi} a_{,\tau}^2 = 1- \frac{2}{M}\left[ \frac{2C^2}{a^4} -\frac{2C^4}{M a^{8}} + \frac{2V_0 C^2}{M a^2} - V_0 a^2  \right]. \ee

\noindent
It is apparent that equation (\ref{Wi}) assures asymptotic flat Euclidean regime as $a_{,\tau}^2 > 0$ and  admits a throat ($a_{,\tau} = 0$) of size
\be a_0 = \left(\frac{2C^2}{M}\right)^{\frac{1}{4}},\ee
which is the same as obtained in the massless case. One can observe that the weak energy condition, ($\rho_{\phi} \ge 0$ and $\rho_{\phi} + p_{\phi} \ge 0$) holds, provided
\be V_0 \le \frac{C^2}{a^6}.\ee
On the other hand, the strong energy condition (which additionally requires $\rho_{\phi} + 3p_{\phi} \ge 0$) holds, provided
\be V_0 \le \frac{4 C^2}{a^6}\left(\frac{ M a^4 -C^2}{5M a^4 - 4C^2}\right).\ee
At the radius of the throat (61) the above conditions (62) and (63) imply
\be V_0 \le \frac{C^2}{ a_{0}^6};\;\;\;\text{and}\;\;\; V_{0} \le \frac{2C^2}{3 a_{0}^6},\ee
giving an upper limit to $V_0$. Hence a viable classical wormhole solution also exists for constant potential.
\subsubsection{Late time Cosmic evolution}
\noindent The gravitational potential along with the back-reaction term given by

\be U_1(a) = \frac{1}{a^2}-\frac{4C^2}{Ma^6} + \frac{4 C^4}{M^2 a^{10}} - \frac{4C^2 V_0}{M^2 a^4}+\frac{2V_0 }{M},\ee
clearly does not suffer from either short distance or long distance instabilities. Further, at the late stage when radiation and matter in the form of pressureless dust are incorporated, the ($^0_0$) equation of Einstein in the flat space ($k = 0$) under wormhole boundary condition, reads,

\be\begin{split} \frac{\dot a^2}{a^2} &= \frac{4}{M}\left[\frac{C^2}{a^6} - \frac{C^4}{M a^{10}} + \frac{C^2 V_0}{M a^4} + \pi^2 \left(\frac{\rho_{ro}}{a^4} + \frac{\rho_{mo}}{a^3}\right) - \frac{V_0}{2} \right] \\
& \approx \frac{4}{M}\left[\Big(\frac{C^2 V_0 + M \pi^2 \rho_{ro}}{M a^4}\Big) + \pi^2 \frac{\rho_{mo}}{a^3} - \frac{V_0}{2} \right],\end{split}\ee
where, we have neglected the first two terms in the last equation as they should be at the late stage of the cosmic evolution. Nevertheless, the constant potential we have started with, behaves as a negative effective cosmological constant under back reaction. As a result, viable late time cosmic evolution is not possible. Thus a scalar field with a constant potential is not suitable for late time cosmological evolution.

\section{Wormholes for power law potentials}
Regular, bounded and well behaved quantum wormhole in the power series approximation and semiclassical wormhole in the WKB approximation had been demonstrated by Hawking and Page \cite{hp:p} for massive scalar field $V = V_0\phi^2$ in the Robertson-Walker minisuperspace model with curvature parameter $k = +1$. For this purpose, they made the approximation $\phi^2 << 1$ and considered large scale factor $a$. Later, Kim \cite{kim} also made a detailed analysis in this connection in the Robertson-Walker metric, for conformally and minimally coupled scalar field both for power law potential of the type $\frac{\lambda_{2p}}{2p}\phi^{2p}$ suitable for chaotic inflationary model and polynomial potential of the type $\frac{\lambda_{2p}}{2p}\phi^{2p} +  2V_0\phi^2$ suitable for new inflationary model, where $p$ is an integer. It was pointed out that operator ordering plays an important role for wave functions to follow Hawking-Page boundary conditions. The solutions were obtained by product integral formulation of wave functions and it was found that half of wave functions were exponentially damped whereas the other half were diverging out at large three geometry. Kim \cite{kim} interpreted the former as tunneling out wave function into and the latter as tunneling in wave function from different universes with the same or different topology. This motivated him to suggest that it is the modulus of wave-function, instead of wave function itself, that should be regular up to some negative power of the three geometry as the three geometry collapses and should be damped at large three geometry. With the help of Liouville-Green transformation, Kim and Page \cite{Kim2} had also shown that wormhole solution for minimally coupled power law scalar field potential exists under the condition that cosmological constant should vanish.
Twamley and Page \cite{Page} also found wormhole solution for minimally coupled imaginary scalar field, taking potential in the forms $V = \frac{1}{4}\lambda\phi^4$ and $V = V_0\phi^2 + \frac{1}{4}\lambda\phi^4$ following Runge-Kutta method of iteration. The solution differs from those obtained by others in the respect that it does not posses conserved charge. It also dispels a conjecture made by Halliwell and Hartle regarding the behaviour of the real part of the action for wormholes possessing complex geometries. It can also overcome the problem with macroscopic wormhole in connection with its stability as argued by Fischler and Susskind \cite{h10}.\\
Inverse power law was first introduced by Peebles and Ratra \cite{pr1}. In recent years potentials with inverse power law had played an important role in explaining late time cosmic acceleration \cite{lt1}. Inverse-power law potential is also linked to particle physics models \cite{pp}. Therefore, in this section we take up power law potential in the form
\be\label{V} V(\phi) = V_{0}\phi^{-\alpha},\ee
where $\alpha$ may be both positive and negative, so that power law potentials can be handled in the same frame. To explore the possibility of obtaining quantum wormholes, following transformation relation
\be\label{N} \eta=a^{m}\phi^{n},\ee
is useful, in view of which  the W-D equation (\ref{WD}) is expressed as
\be\label{E}
\frac{\hbar^2}{2M}\left(a^2 \Psi_{aa}+ap\Psi_{a}\right) - \frac{M}{2}ka^4 \Psi =\frac{\eta^{\frac{2n-2}{n}}}{a^{-\frac{2m}{n}}}\left[\frac{\hbar^2}{2}\left(n^{2}
\Psi_{\eta\eta}+\frac{n(n-1)}{\eta}\Psi_{\eta}\right)-V_{0} a^{\left(6+\frac{m(\alpha-2)}{n}
\right)} \eta^{\left(\frac{2-2n-\alpha}{n}\right)} \Psi\right].
\ee
Separation of variable in the form $\Psi(a,\eta) = A(a)B(\eta)$ is possible, provided
\be\label{M} m = \frac{6n}{2-\alpha}\;\; \text{and} \;\;\alpha \ne 2\ee
for finite value of $m$ and for a meaningful form of the wave function as well. However, condition (\ref{M}) implies, that inverse square law potentials $(V(\phi) = V_0 \phi^{-2})$ can not be treated in the same frame. Thus we have,
\be\label{PP}
\frac{a^{-\frac{2m}{n}}}{A}\left[\frac{\hbar^2}{2M}\left(a^2 A_{aa}+ap A_{a}\right)-
\frac{M}{2}ka^4 A\right] =\frac{\eta^{\frac{2n-2}{n}}}{B}\left[\frac{\hbar^2}{2}\left(n^{2}~
B_{\eta\eta}+\frac{n(n-1)}{\eta}~B_{\eta}\right) -V_{0}~  \eta^{\frac{2(3-m)}{m}}~B\right]
=\omega^2,
\ee
where, $\omega^2$ is the separation constant.

\subsection{Massive scalar field $ V =V_{0}\phi^2 $ }
\subsubsection{Quantum Wormhole}
\noindent
This case corresponds to $\alpha=-2$ in view of (\ref{V}). Now if we choose $n=1$ then equation (\ref{M}) yields $m = 3/2$, implying $\eta = a^{3/2}\phi$ in view of (\ref{N}). In the process, we find that the solution of (\ref{PP}) exists only under the choice for the separation constant $\omega^2 = 0$. The solutions are
\be\label{AA} A(a)=a^{\frac{1-p}{2}}\left(\frac{2^4 \hbar^2}{k M^2}\right)^\frac{p-1}{8} \left[C_{7}(-1)^\frac{1-p}{4} I_{\big(\frac{1-p}{4}\big)}\left(\frac{a^2\sqrt k M}{2\hbar}\right)\Gamma\left(\frac{5-p}{4}\right) + C_{8}I_{-\big(\frac{1-p}{4}\big)}\left(\frac{a^2\sqrt k M}{2\hbar}\right)\Gamma\left(\frac{3+p}{4}\right)\right] \ee
and
\be B(\eta)= C_9 {D}_{\left(-\frac{1}{2}\right)}\left(\frac{(8 V_{0})^\frac{1}{4}~ \eta}{\sqrt \hbar}\right) + C_{10}{D}_{\left(-\frac{1}{2}\right)}\left(i\frac{(8 V_{0})^\frac{1}{4}~ \eta}{\sqrt \hbar}\right) \ee

\noindent
Here, $I_\alpha(x)$, ${D_\nu(x)}$ are the modified Bessel function of first kind and Parabolic cylinder function respectively while, $C_{7}, C_{8}, C_{9}$ and $C_{10}$ are integration constants. The real part of the wavefunction $ \Psi(a,\eta) = A(a)B(\eta) $ exhibits ultraviolet divergence for $p > 1$. Nevertheless, for $p \leq 1$ the Parabolic cylinder function controls the divergence (see appendix) appearing in modified Bessel function and hence $\Psi(a,\eta)$ becomes regular at both ends. Thus wormhole exists for $p \le 1$, which has been exhibited in figure 1, setting $C_7 = C_8 = C_{9} = \hbar = M = k = p = 1$.
\begin{figure}
[ptb]
\begin{center}
\includegraphics[
height=2.034in, width=2.8in] {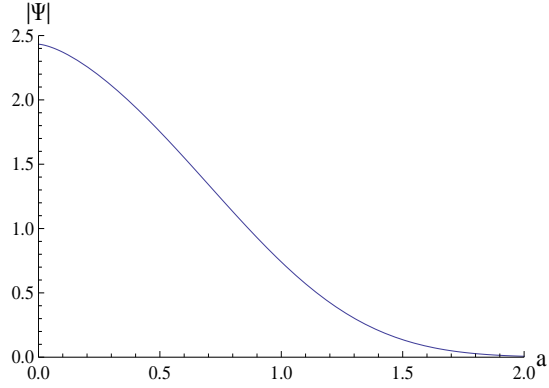} \caption{The figure depicts that $|\Psi|$ satisfies Hawking-Page wormhole boundary condition in the case of $V = V_0 \phi^2$ for $p \le 1$ and $\phi > 0$. The figure has been plotted setting the constants $C_7 = C_8 = C_{9} = \hbar = M = k = p = \phi = 1$ after testing a wide number of cases with different values of the constants viz., $C_7$, $C_8$, $C_{9}$ and $p$ for which the nature of $|\Psi|$ remains unaltered.}
\end{center}
\end{figure}

\subsubsection{Semiclassical Wormhole}

Equation (\ref{7}) now takes the form

\be \pm 2\hbar\sqrt k\frac{\partial f}{\partial a} = -\frac{\hbar^2}{a^3} \frac{\partial^2 f}{\partial\phi^2}+2V_0 a^3\phi^2 f. \ee

\noindent Here again let us choose the same variable, $\eta = a^{\frac{3}{2}}\phi$, to reduce the above equation to

\be \pm 2\hbar\sqrt k \frac{1}{f}\left(\frac{\partial f}{\partial a}\right)  = -\hbar^2 \frac{1}{f}\left(\frac{\partial^2 f}{\partial\eta^2}\right)+2V_0 \eta^2. \ee

\noindent Separating the variables as $f(a, \eta) =  A(a)B(\eta)$, we have

\be \pm2\hbar\sqrt k \frac{1}{A}\left(\frac{\partial A}{\partial a}\right) = -\hbar^2\frac{1}{B}\left(\frac{\partial^2 B}{\partial\eta^2}\right)+2V_0\eta^2 = \omega^2. \ee

\noindent where, $ \omega^2 $ is the separation constant. Solving the above equations for $ a $ and $ \eta $ we get,

\be f(a, \eta) = A_0 \exp{\left( \pm \frac{ \omega_0^2 \hbar a}{2\sqrt k}\right)} \times \left[ C_{11}{D}_{\frac{\omega_0^2 - \omega_1}{2\omega_1}}\left(\sqrt {2\omega_1}\eta\right) + C_{12}{D}_{\left(-\frac{\omega_0^2 + \omega_1}{2\omega_1}\right)}\left(i\sqrt {2\omega_1}\eta\right) \right], \ee

\noindent where,
\be\label{11} \omega_0^2 = \frac{\omega^2}{\hbar^2} \hspace{.2 in} \mathrm{and} \hspace{ .2 in}  \omega_1^2 =2\frac{V_0}{\hbar^2}. \ee
Here again ${D_\nu(x)}$ denotes Parabolic Cylinder function, while $C_{11}$, $C_{12}$ are constants. Now taking only real part of the above equation and in view of (\ref{6}), we get the following form of the wavefunction, viz.,

\be\label{P2} \Psi(a,\phi) = (A_0 C_{11}\mu) a^{-\frac{p+1}{2}}\exp{\left[-\frac{1}{\hbar}\left(\frac{M}{2}\sqrt{k}a^2 - \frac{\omega_0^2 \hbar^2 a}{2\sqrt k}\right)\right]} \times {D}_{\frac{\omega_0^2 - \omega_1}{2\omega_1}}\left(\sqrt {2\omega_1}\eta\right).\ee

\noindent
Since parabolic cylinder function (see appendix) is well behaved at both ends ($a \rightarrow 0$ and $a \rightarrow \infty$), therefore $\Psi$ satisfies wormhole boundary condition for $p \le -1$ for both positive and negative sign in the exponential for $k = +1$. Note that semiclassical wormhole is much restrictive than the quantum one since quantum wormhole is admissible for $p \le 1$. Here we would like to mention that Hawking and Page set $\phi^2 << 1$ at $a \rightarrow 0$, to satisfy W-H boundary condition for all values of $p$ \cite{hp:p}. \\

\subsubsection{Back-reaction and the throat}

To find the wormhole throat we proceed as before (ie., express the exponent of the solution (\ref{P2}) as $\exp[\frac{i}{\hbar}S_t]$) to obtain the Einstein's ($^0_0$) equation with back-reaction term as,

\be\label{14} \frac{\dot a^2}{a^2}+\frac{1}{a^2}= \frac{2}{M}<\left(\frac{1}{2}\dot\phi^2 + V_0\phi^2\right)> = \frac{2}{M}<\rho_{\phi}>  =\frac{2}{M}\left( \frac{\omega_0^2\hbar^2}{2a^3}-\frac{\omega_0^4\hbar^4}{8 Ma^4} \right),\ee
which under Wick rotation reads
\be a_{,\tau}^2 = 1- \frac{2}{M}\left( \frac{\omega_0^2\hbar^2}{2a}-\frac{\omega_0^4\hbar^4}{8 Ma^2} \right).\ee
$a_{,\tau}^2$  is clearly positive assuring asymptotic flat Euclidean universe and the radius of the throat of the wormhole is,
\be a_0 = \frac{\omega_0^2\hbar^2}{2M}, \ee
Now the strong energy condition requires
\be \rho_{\phi}+ p_{\phi} = \frac{\omega_0^2\hbar^2}{a^3} - \frac{\omega_0^4\hbar^4}{4 Ma^4} - 2 V_0\phi^2 > 0.\ee
\be \rho_{\phi}+3 p_{\phi} = \frac{2\omega_0^2\hbar^2}{a^3} - \frac{\omega_0^4\hbar^4}{2 Ma^4} - 6 V_0\phi^2 > 0.\ee
Therefore at the throat, the strong energy condition is satisfied under the condition $V(\phi) \le \frac{4M^3}{3 {\omega_0}^4 {\hbar}^4}$. The condition that the potential should not be too large at any stage is of-course justifiable. Hence a well behaved classical wormhole is obtainable from the semiclassical one, under back-reaction.
\subsubsection{Late time Cosmic evolution}
The gravitational potential
\be U_2(a) =  \frac{1}{a^2}-\frac{\omega_0^2\hbar^2}{Ma^3}+\frac{\omega_0^4\hbar^4}{4 M^2a^4}\ee
here again does not suffer from any short or long distance instability. As discussed earlier, if we add some contribution from radiation and matter in the form of dust as well (for $k = 0$), the ($^0_0$) component of Einstein's equation together with the wormhole boundary condition reads,
\be \frac{\dot a^2}{a^2} = \frac{2}{M}\left[ \frac{1}{a^4}\Big(2\pi^2 \rho_{r0}-\frac{\omega_0^4\hbar^4}{8 M}\Big) + \frac{1}{a^3}\Big(2\pi^2\rho_{m0} + \frac{\omega_0^2\hbar^2}{2}\Big)\right].\ee
Thus the back-reaction terms only reduces the radiation density and increases the matter density slightly and hence keeps the Friedmann solutions ($a \propto \sqrt t$ in radiation era and $a \propto t^{\frac{2}{3}}$ in the matter dominated era) along with all cosmological observations unaltered. This is of-course a wonderful feature of massive scalar field since late time cosmological evolution remains unaltered from the standard Friedmann cosmology. Thus massive scalar field $V = V_0 \phi^2$ is the best candidate to describe the history of cosmological evolution.

\subsection{Quartic potential, $V(\phi) = \lambda \phi^4$.}

\subsubsection{Quantum wormhole}

\noindent This case in view of equation (\ref{V}) corresponds to $\alpha= - 4$. Now the choice $m = n = 1$ clearly satisfies equation (\ref{M}) which results in $\eta = a \phi$ in view of equation (\ref{N}). Equation (\ref{PP}) may now be solved again only for $\omega^2 = 0$. The solution of $A(a)$ is therefore the same modified Bessel function of first kind presented in (\ref{AA}), while $B(\eta)$ may be solved as

\begin{equation}
B(\eta)=\left(\frac{\lambda}{18 \hbar^2}\right)^\frac{1}{12}\sqrt \eta \left[C_{13}(-1)^\frac{1}{6}I_{\frac{1}{6}}\left(\frac{\sqrt {2\lambda}~
\eta^3}{3\hbar}\right)\Gamma\left(\frac{7}{6}\right)+C_{14}I_{-\frac{1}{6}}\left(\frac{\sqrt {2\lambda}~
\eta^3}{3\hbar}\right)\Gamma\left(\frac{5}{6}\right)\right]
\end{equation}

\noindent where, $C_{13}$ and $C_{14}$ are integration constants. The wavefunction, $\Psi(a,\eta) = A(a)B(\eta)$ is a product of two modified Bessel functions [$I_\alpha(x)$] and so suffers from both infrared and ultraviolet divergences for $p > 1$, while it exhibits infrared divergence for $p \le 1$. Thus quantum wormhole does not exist for quartic potential $V = \lambda \phi^4$.\\

\subsubsection{Semiclassical Wormhole}
\noindent In this case equation (\ref{7}) takes the form

\be \pm 2\hbar\sqrt k\frac{\partial f}{\partial a} = -\frac{\hbar^2}{a^3} \frac{\partial^2 f}{\partial\phi^2}+2 \lambda a^3\phi^4 f. \ee

\noindent Choosing a new variable $\eta = a \phi $ as in the quantum case, the above equation can be rewritten as

\be \pm 2 \hbar\sqrt k ~\frac{a}{f}\left(\frac{\partial f}{\partial a}\right) = -\hbar^2\frac{1}{f}\left(\frac{\partial^2 f}{\partial \eta^2}\right)+ 2 \lambda \eta^4. \ee

\noindent
Now using the method of separation of variables by taking $f(a, \eta) =  A(a)B(\eta)$, we have

\be \pm 2 \hbar\sqrt k ~\frac{a}{A}\left(\frac{\partial A}{\partial a}\right) = -\hbar^2\frac{1}{B}\left(\frac{\partial^2 B}{\partial \eta^2}\right)+ 2 \lambda \eta^4 = \omega^2, \ee

\noindent where $\omega^2$ is the separation constants. The solutions of the above equation exists here again, only for $\omega^2=0$, which are

\be A(a)=constant = A_0 \ee
\be B(\eta)=\sqrt \eta \left(\frac{\lambda}{18\hbar^2}\right)^{\frac{1}{12}}\left[C_{15}(-1)^\frac{1}{6}
I_{\frac{1}{6}}\left(\frac{\sqrt{2 \lambda}\eta^3}{3\hbar}\right)\Gamma\left(\frac{7}{6}\right)+C_{16}I_{-\frac{1}{6}}\left(\frac{\sqrt{2 \lambda}\eta^3}{3\hbar}\right)\Gamma\left(\frac{5}{6}\right)\right] \ee

\noindent
where, as already mentioned, $I_\alpha(x)$ is the modified Bessel function of first kind and $C_{15}$ and $C_{16}$ are integration constants. Now in view of equation (\ref{6}) the wavefunction takes the following form,
\be \Psi(a,\eta) =  A_0 \mu a^{-\frac{p+1}{2}}\exp{\left(-\frac{M}{2\hbar}\sqrt k a^2\right)}B(\eta). \ee

\noindent
This wave function clearly does not satisfy wormhole boundary condition and so semiclassical wormhole also does not exist for quartic potential.

\subsubsection{Series solution}

No one has obtained wormhole configuration for quartic potential in straight forward manner, as discussed at the beginning of this section. We therefore make yet another attempt to find series solution for equation (\ref{7}) under the choice $f = f(\eta)$, where, $\eta = a \phi$, as before. Equation (\ref{7}) may then be expressed as,

\be \label{16} \frac{d^2 f}{d\eta^2}\pm\left(\frac{2\eta\sqrt k}{\hbar}\right)\frac{d f}{d\eta}-\left(\frac{2\lambda \eta^4}{\hbar^2}\right)f=0.\ee

\noindent
Further, under the choice
\be f = g(\eta)\exp{\left(-\frac{b\eta^2}{4}\right)}, \hspace{ .2 in} \text{where,} \hspace{ .2 in} b = \mp\frac{\sqrt k}{\hbar}, \ee

\noindent
the above equation (\ref{16}) may be reexpressed as,
\be \frac{d^2g}{d\eta^2}+\left(\mp\frac{\sqrt k}{\hbar}+\frac{k}{\hbar^2}\eta^2 -\frac{2\lambda}{\hbar^2}\eta^4\right)g=0.\ee

\noindent
Now taking,
\be g = \sum^{\infty}_{n=0} g_n\eta^n \ee
equation (96) reads

\be -\frac{2\lambda}{\hbar^2}g_n+\frac{k}{\hbar^2}g_{n+2}\mp\frac{\sqrt k}{\hbar}g_{n+4}+(n+5)(n+6)g_{n+6}=0.\ee

\noindent
All the coefficients from $g_2$ to $g_6$ can now be found in terms of $g_0$ and $g_1$ which remain arbitrary, as follows, .

\begin{eqnarray}
  g_2 &=& \pm\frac{\sqrt k}{2\hbar}g_0 \nonumber \\
  g_3 &=& \pm\frac{\sqrt k}{2.3\hbar}g_1 \nonumber \\
  g_4 &=& -\frac{6g_0}{2.3.4}-\frac{k}{2\hbar^2}\frac{g_3}{2.3.4} \nonumber \\
  g_5 &=& \pm \frac{\sqrt k}{(2.3.4.5)\hbar} g_3 - \frac{k}{(2.3.4.5)\hbar^2} g_1 \nonumber \\
  g_6 &=& \pm \frac{\sqrt k}{(2.3.4.5.6)\hbar}g_4 - \frac{k}{(2.3.4.5.6)\hbar^2}g_2 + \frac{2\lambda}{(2.3.4.5.6)\hbar^2}g_{0}.
\end{eqnarray}

\noindent
Thus the solution to $f(a,\phi)$ is

\be\label{SS} f(a,\phi) = \sum _{n=0}^{\infty}g_n (a\phi)^n\exp{\left(\pm\frac{\sqrt k}{\hbar}\times\frac{a^2\phi^2}{4}\right)}.\ee
So in view of (\ref{6}) the  wavefunction finally takes the following form,
\be \Psi(a,\phi) =  \mu a^{-\frac{p+1}{2}}\times \left( \sum _{n=0}^{\infty}g_n (a\phi)^n \right)\times \exp{\left(-\frac{\sqrt k a^2}{2\hbar}\left(M \mp \frac{1}{2} \right) \right)}\times \exp{\left( \pm \frac{\sqrt k}{4\hbar}\phi^2 \right) }. \ee

\noindent
Note that both the terms in the exponent are now having the same form and it appears that for $k = +1$ and $p \leq -1$ the wave function is well behaved at both ends. Nevertheless, at small three volume the wave function vanishes irrespective of the signature (instead of being finite) as has been depicted in figure 2. As a result the formation of baby universe with a finite throat is not possible.

\begin{figure}
[ptb]
\begin{center}
\includegraphics[
height=2.0in, width=2.9in] {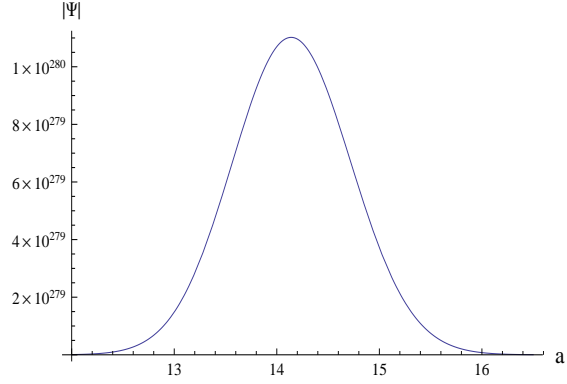} \caption{Interestingly enough the series solution for $V = \lambda\phi^4$ shows gaussian nature for large but finite $n$ irrespective of the sign appearing in the exponent. Since the wavefunction vanishes as $a\rightarrow 0$, so formation of baby universe with a finite throat remains obscure even for $p \le -1$. This graph has been plotted with $k = \hbar = M = 1, p = -1$ and taking the first $n = 300$ terms in the series solution. For larger and larger $n$, the maxima increases and the graph shifts to the right.}
\end{center}
\end{figure}

\subsection{$V = \frac{V_{0}}{\phi}$}

\subsubsection{Quantum Wormhole}

Inverse potentials, as mentioned has been found proved its importance in the context of late time acceleration. So, to study the effect of such potentials in the early universe, at first let us take it as $V(\phi) = \frac{V_0}{\phi}$, which in comparison with equation (\ref{V}) requires $\alpha=1$. Further choosing $n=1, m=6$, so that $\eta = \phi a^6$ in view of equation (\ref{N}), the equation for $A(a)$ reads,
\be\label{39} \frac{\hbar^2}{2M}\left(A_{aa}+\frac{p}{a}A_{a}\right)-\frac{M}{2}ka^2 A-\omega^{2} a^{10}A = 0 \ee

\noindent
Equation (\ref{39}) can be solved under the choice $\omega^2 = 0$, resulting in the same solution as in (\ref{AA}). In general, for arbitrary $\omega^2$, series solution of equation (\ref{39}) can be found which shows a regular singularity of pole of order one and one can find indicial roots for it. Further the $B(\eta)$ equation
\be \frac{\hbar^2}{2}B_{\eta\eta}- \frac{V_{0}}{\eta}B-\omega^2 B = 0,\ee

\noindent
under the same choice $\omega^2 = 0$, yields the solution
\be B(\eta)=\frac{\sqrt{2\eta V_0}}{\hbar} \left[-C_{17} I_1\left(\frac{\sqrt{8 \eta V_{0}}}{\hbar} \right) + 2C_{18} K_1\left(\frac{\sqrt{8 \eta V_{0}}}{\hbar} \right)\right] \ee

\noindent
where, $I_\alpha(x)$ and $K_\alpha(x)$ are modified Bessel functions of first and second kind respectively, while $C_{17}$ and $C_{18}$ are arbitrary constants. The wave function [$\Psi(a, \eta) = A(a)B(\eta)$] shows both UV and IR divergences for $p > 1$ and IR divergence for $p \le 1$. However, if we consider $K_\alpha(x)$ to be the particular solution of $B(\eta)$ as

\be B(\eta)=\frac{\sqrt{8\eta V_0}}{\hbar} K_1\left(\frac{\sqrt{8 \eta V_{0}}}{\hbar} \right), \ee

\noindent then wormhole boundary condition is satisfied for $p\le1$ (see appendix). The plot of such regular wave function $|\Psi| = |A(a)B(\eta)|$ has been presented in figure-3, setting $k = M = C_{17} = C_{18} = \phi = V_0 = \hbar = p = 1$.

\begin{figure}
\begin{center}
\includegraphics[
height=2.0in, width=2.9in] {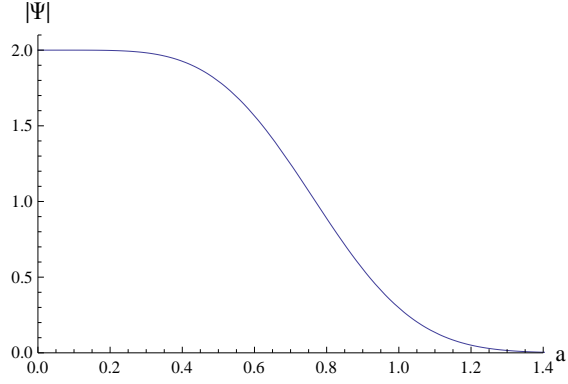} \caption{This is the plot of quantum solution of $|\Psi|$ for the case $V(\phi)=\frac{V_0}{\phi}$, where we have set $k = M = C_{17} = C_{18} = \phi = V_0 = \hbar = p = 1$. It is important to mention that other choices of constants do not alter the shape of the graph.}
\end{center}
\end{figure}

\subsubsection{Semiclassical wormhole}

Under the same above choice viz, $\eta = \phi~ a^6$ with $n = 1$ and $m = 6$ equation (\ref{7}) can be expressed as,
\be \pm 2\frac{\hbar\sqrt k}{a^9} \frac{1}{f}\left(\frac{\partial f}{\partial a}\right) = -\hbar^2\frac{1}{f}\left(\frac{\partial^2 f}{\partial \eta^2}\right)+\frac{2V_0}{\eta}. \ee

\noindent
Now separation of variables in the form $f(a,\eta) = A(a)B(\eta)$ leads to
\be \pm 2\frac{\hbar\sqrt k}{a^9} \frac{1}{A}\left(\frac{\partial A}{\partial a}\right) = -\hbar^2\frac{1}{B}\left(\frac{\partial^2 B}{\partial \eta^2}\right)+\frac{2V_0}{\eta} = -\omega^2. \ee

\noindent
Equation (107) may now be solved to yield,
\be A(a) = A_0\exp{\left(\mp \frac{\omega^2a^{10}}{20\hbar\sqrt k}\right)} \ee
and
\be B(\eta)= \frac{1}{\hbar^2} \eta e^{- \omega_{0} \eta}\left[C_{19}\; {_1F_1}\left(1+ \frac{\omega_{1}^2}{2\omega_{0}},2;2 \omega_{0} \eta \right)+C_{20}\; U\left(1+ \frac{\omega_{1}^2}{2\omega_{0}},2;2 \omega_{0} \eta \right)\right] \ee

\noindent
where, $\omega_0$ and $\omega_1$ are the same as given in equation (\ref{11}) while, $A_0$, $C_{19}$ and $C_{20}$ are integration constants. ${_1F_1}(a,b;x)$ and $U(a,b;x)$ are confluent hypergeometric functions of first and second kind respectively. So in view of equation (\ref{6}) the wavefunction for this particular case is found as,

\be\label{18} \Psi(a,\eta) =  \mu a^{-\frac{p+1}{2}}\exp{\left[-\frac{1}{\hbar}\left(\frac{M}{2}\sqrt k a^2 + \frac{\omega^2a^{10}}{20\sqrt k}\right) \right]}B(\eta).\ee

\noindent Under the Choice $C_{19}=0$ the wave function $\Psi(a,\phi)$ admits wormhole boundary condition for $p \leq -1$ and $k = +1$ (see appendix).

\subsubsection{Back-reaction and throat}

As before, setting $k = +1$, the Einstein's ($^0_0$) equation with back-reaction terms takes the form,
\be \label{26}\frac{\dot a^2}{a^2}+\frac{1}{a^2}= -\frac{2}{M}\left(\frac{\omega^2 a^6}{ 2} + \frac{\omega^4 a^{14}}{8M} \right), \ee
which under Wick rotation reads
\be a_{,\tau}^2 = 1+\frac{2}{M}\left(\frac{\omega^2 a^8}{ 2} + \frac{\omega^4 a^{16}}{8M} \right).\ee
Although $a_{,\tau}^2 > 0$, the throat clearly becomes imaginary. Thus semiclassical wormhole although became apparent from the wavefunction (\ref{18}), its classical counterpart does not yield a viable wormhole solution. Further, the effective gravitational potential shows infrared divergence which implies inflation never halts. Since wormhole exists in the quantum domain therefore, such inverse form of potential is nice to explain very early universe but is not suitable to explain late time cosmic evolution.

\subsection{$V = \frac{V_{0}}{\phi^{3}}$}
\subsubsection{Quantum Wormhole}

\noindent In comparison with equations (\ref{V}), (\ref{N}) and (\ref{M}) the present case is found to correspond with $\alpha = 3, n = 1$ and $m = -6$, so that $\eta = \frac{\phi}{a^6}$. As before, the solution is found in a straight forward manner choosing the separation constant $\omega^2 = 0$. Obviously, the solution $A(a)$ is same as in equation (\ref{AA}) and the solution for $B(\eta)$ is given below

\be \label{phi-3} B(\eta)=\hbar \sqrt{\frac{\eta}{2 V_{0}}} \left[C_{21}{I}_1\left(\frac{2}{\hbar} \sqrt{\frac{2 V_{0}} {\eta}} \right)+2 C_{22}{K}_1\left(\frac{2}{\hbar} \sqrt{\frac{2 V_{0}} {\eta}} \right)\right] \ee

\noindent where, $I_{\alpha}(x)$ and $K_{\alpha}(x)$ are modified Bessel functions as already mentioned while $C_{21}$ and $C_{22}$ are constants of integration. In this case the wavefunction $\Psi = A(a) B(\eta)$ exhibits both UV and IR divergences for all $p$ as depicted in figure 4 and so quantum wormhole does not exist. Wormhole boundary condition is not satisfied even under the choice $C_{21} = 0$, due to the appearance of a multiplicative factor $\sqrt \eta = \frac{\phi}{a^3}$ (see appendix).\\

\begin{figure}
[ptb]
\begin{center}
\includegraphics[
height=2.034in, width=2.8in] {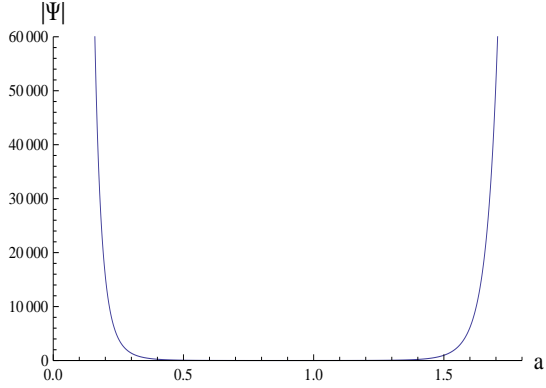} \caption{Quantum wormhole wavefunction $|\Psi|$ shows both UV and IR divergence for all values of $p$ in the case of inverse potential in the form $\frac{V_0}{\phi^3}$. The present plot has been presented setting $C_{21}= C_{22} = M = V_0 =\phi= \hbar= p= k= 1$.}
\end{center}
\end{figure}

\subsubsection{Semiclassical Wormhole}
\noindent In this case equation (\ref{7}) takes the form

\be \pm 2\hbar\sqrt k\frac{\partial f}{\partial a} = -\frac{\hbar^2}{a^3} \frac{\partial^2 f}{\partial\phi^2}+2 V_0 \frac{a^3}{\phi^3} f. \ee

\noindent Choosing the same new variable $\eta = \frac{\phi}{a^6}$ as in the quantum case, the above equation can be rewritten as

\be \pm 2 \hbar\sqrt k ~\frac{a^{15}}{f}\left(\frac{\partial f}{\partial a}\right) = -\hbar^2\frac{1}{f}\left(\frac{\partial^2 f}{\partial \eta^2}\right)+ 2 \frac{V_0}{\eta^3}. \ee

\noindent
Now using the method of separation of variables by taking $f(a, \eta) =  A(a)B(\eta)$, we have

\be \pm 2 \hbar\sqrt k ~\frac{a^{15}}{A}\left(\frac{\partial A}{\partial a}\right) = -\hbar^2\frac{1}{B}\left(\frac{\partial^2 B}{\partial \eta^2}\right)+ 2 \frac{V_0}{\eta^3} = \omega^2, \ee

\noindent where $\omega^2$ is the separation constants. The solutions of the above equation here again exists only for $\omega^2=0$. The solutions are $A(a) = constant = A_0$ while $B(\eta)$ is the same modified Bessel function as presented in equation (\ref{phi-3}). Now in view of equation (\ref{6}) the wavefunction takes the following form,
\be \Psi(a,\eta) =  A_0 \mu a^{-\frac{p+1}{2}}\exp{\left(-\frac{M}{2\hbar}\sqrt k a^2\right)}B(\eta).\ee

\noindent
Since the modified Bessel function $B(\eta)$ itself shows divergence in its behaviour, so the Van-Vleck determinant $(\mathcal{D})$ also remains unregulated and the wave function does not satisfy wormhole boundary condition. Thus, semiclassical wormhole remains absent also for $V =  \frac{V_0}{\phi^3}$.

\subsection{$ V = \frac{V_{0}}{\phi^{4}}$}
\subsubsection{Quantum Wormhole}

\noindent This case correspond to, $\alpha = 4 , n = 1, m = -3$ with $\eta = \frac{\phi}{a^3}$ as can be seen while comparing with equations (\ref{V}), (\ref{N}) and (\ref{M}). As before, solution is found in a straight forward manner only under the choice $\omega^2 = 0$. Obviously, the solution for $A(a)$ is again the same as in equation (\ref{AA}) and the solution $B(\eta)$ is

\be \label{phi-4} B(\eta)=\eta \left(C_{23}~e^{\frac{\sqrt {2 V_{0}}}{\eta \hbar}} + C_{24} \frac{\hbar}{\sqrt {8 V_{0}}}~e^{-\frac{\sqrt {2 V_{0}}}{\eta \hbar}}\right) \ee

\noindent where $C_{23}$ and $C_{24}$ are the constants of integration. Clearly the first term shows divergence in its behaviour while the second term is well behaved at both ends. Therefore the solution of $B(\eta)$ in no way can control the diverging behaviour of the solution for $A(a)$ presented in (\ref{AA}). Hence the wavefunction $\Psi$ exhibits both UV and IR divergences for all values of $p$ in the similar fashion as depicted in figure 4.

\subsubsection{Semiclassical Wormhole}
\noindent In this case equation (\ref{7}) takes the form

\be \pm 2\hbar\sqrt k\frac{\partial f}{\partial a} = -\frac{\hbar^2}{a^3} \frac{\partial^2 f}{\partial\phi^2}+2 V_0 \frac{a^3}{\phi^4} f. \ee

\noindent Choosing the same new variable $\eta = \frac{\phi}{a^3} $ as taken in the quantum case, the above equation can be rewritten as

\be \pm 2 \hbar\sqrt k ~\frac{a^9}{f}\left(\frac{\partial f}{\partial a}\right) = -\hbar^2\frac{1}{f}\left(\frac{\partial^2 f}{\partial \eta^2}\right)+ 2 \frac{V_0}{\eta^4}. \ee

\noindent
Now using the method of separation of variables under the choice $f(a, \eta) =  A(a)B(\eta)$, we have

\be \pm 2 \hbar\sqrt k ~\frac{a^9}{A}\left(\frac{\partial A}{\partial a}\right) = -\hbar^2\frac{1}{B}\left(\frac{\partial^2 B}{\partial \eta^2}\right)+ 2 \frac{V_0}{\eta^4} = \omega^2, \ee

\noindent where $\omega^2$ is the separation constants. The above differential equation for $B(\eta)$ admits solution in closed form only when $\omega^2=0$, which is already present in the equation (\ref{phi-4}), whence $A(a)=constant = A_0$. Thus in view of equation (\ref{6}) the wavefunction takes the following form,
\be \Psi(a,\eta) =  \frac{C_{24} A_0 \mu \hbar}{\sqrt {8V_0}} a^{-\frac{p+7}{2}}\exp{\left[-\frac{1}{\hbar}\Big(\frac{M}{2}\sqrt k a^2 + \frac{\sqrt{2V_0}}{\phi} a^3\Big)\right]}\phi.\ee
\noindent
Here we have taken only the second part of the solution (\ref{phi-4}) setting $C_{23} = 0$. The wave function (122) is found to satisfy wormhole boundary condition for $p \le -7$, $k = +1$ and for regular functional behaviour of $\phi$.

\subsubsection{Back-reaction and the throat}
To find the back-reaction term we choose
\be \frac{i}{\hbar}S_t = -\frac{M}{2\hbar}\sqrt k a^2 - \frac{2V_0}{\hbar \phi} a^3\ee
and identify $S_{t,a}$ with the classical momentum as before, to obtain

\be\frac{\dot a^2}{a^2} + \frac{1}{a^2} = - \frac{2}{M}\left(\frac{18 V_0^2}{M\phi^2} + \frac{6 V_0}{ \phi a}\right),\ee
which under Wick rotation reads
\be a_{,\tau}^2 = 1 + \frac{2}{M}\left(\frac{18 V_0^2 a^2}{M\phi^2} + \frac{6 V_0 a}{ \phi}\right),\ee
Here again we encounter the same situation as occurred in the case of inverse potential in the form $V_{0}/\phi$, ie., although $a_{,\tau}^2 > 0$, the throat does not have a real root. Thus semiclassical wormhole (122) does not yield a classical counterpart. Nevertheless, unlike the previous situation referred, the effective gravitational potential

\be U(a) = \frac{1}{a^2} + \frac{2}{M}\left(\frac{18 V_0^2}{M\phi^2} + \frac{6 V_0}{ \phi a}\right),\ee
does not show infrared divergence.

\section{Quantum and semiclassical Wormholes for Exponential potential.}

\subsection{Exponential potential in the form $V = V_{0} e^{-\phi/\lambda}$}

\subsubsection{Quantum Wormhole}

Exponential potential in the said form was first introduced by Ratra and Peebles \cite{rp}. Later it was found to play a significant role to explain late time cosmic acceleration \cite{lt}. However, here we consider both the signs of $\lambda$. To solve the W-D equation (\ref{WD}) for the case under consideration, let us make a change of variable as $y = a^6 e^{-\phi/\lambda}$. Now separating the wave function as $ \Psi = A(a)B(y) $ and taking $\omega^2$ as separation constant, the W-D equation can be expressed as
\be\label{19} \frac{\hbar ^2}{2M}\left(a^2 A_{aa} + p a A_{a}\right)- \left(\frac{M}{2}ka^4 + \omega^2\right)A = 0, \ee
and
\be\label{20} \frac{\hbar ^2}{2\lambda^2}\left(y^2 B_{yy}+y B_{y}\right)-\left(V_{0} y + \omega^2\right)B = 0 . \ee

\noindent
It is to be noted that $\lambda^2$ appears in equation (\ref{20}) and so the solution is independent of the choice of the signature of $\lambda$. The solutions of equation (\ref{19}) is,

\be A(a)=(-1)^{\frac{1-p-4x}{8}} \left(\frac{2^4 \hbar^2}{a^4 k M^2}\right)^\frac{p-1}{8} \left[C_{25}(-1)^x I_x\left(\frac{a^2\sqrt k M}{2\hbar}\right)\Gamma\left(1+x \right)+C_{26}I_{-x}\left(\frac{a^2\sqrt k M}{2\hbar}\right)\Gamma\left(1-x \right)\right] \ee
where, $C_{25}$ and $C_{26}$ are constants of integration and
\be x=\frac{\sqrt{\hbar^2 (p-1)^2 + 8 M \omega^2}}{4 \hbar}.\ee

\noindent
The solution of equation (\ref{20}) can be written in the following form,
\be B(y)=(-1)^{\frac{x_{1}}{2}}\left[C_{27}I_{x_1}\left(x_{2}\right)\Gamma \left(1+x_{1} \right)+C_{28}I_{-x_1}\left(x_{2}\right) \Gamma \left(1-x_{1} \right)\right] \ee

\noindent
$C_{27}$ and $C_{28}$ being the constants of integration while, $I_{\alpha}(x)$ is the modified Bessel function of the first kind, with $x_{1}=\frac{2\sqrt 2 \omega \lambda}{\hbar}$ and $x_{2}=\frac{\sqrt {8 y V_{0}}\lambda}{\hbar}$. Here again, the wavefunction $\Psi$ shows both the UV and IR divergences for all values of $p$ as presented in figure 5.

\begin{figure}
[ptb]
\begin{center}
\includegraphics[
height=2.034in, width=2.8in] {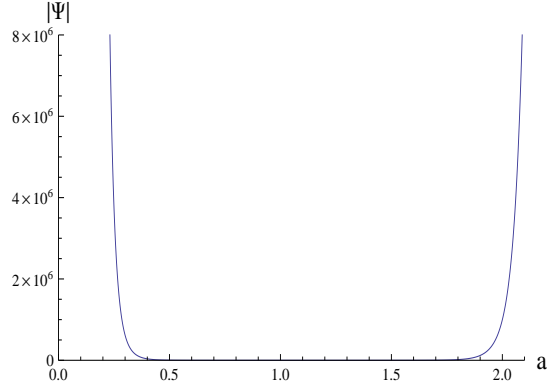} \caption{$|\Psi|$ shows both UV and IR divergence behavior for all $p$ in case of exponential potential. In the present plot we have chosen $C_{25} = C_{26} = C_{27} = C_{28} =\hbar = V_0 = M = p = k =\omega = \lambda = \phi = 1$}
\end{center}
\end{figure}

\subsubsection{Semiclassical Wormhole}

Expressing equation (\ref{7}) in the following form,
\be\label{42} \pm \hbar\sqrt k\frac{a^3}{f}\frac{\partial f}{\partial a} = -\frac{\hbar^2}{2}\frac{1}{f}\frac{\partial^2 f}{\partial\phi^2} + a^6e^{-\frac{\phi}{\lambda}}, \ee

\noindent
it is apparent that the above equation is not separable in its present form. So let us consider a change of variable,
\be \alpha = \ln{(a^6e^{-\frac{\phi}{\lambda}})} = 6\ln a - \frac{\phi}{\lambda} \ee

\noindent
and equation (\ref{42}) may  now be expressed as,
\be \pm \hbar\sqrt k\frac{a^3}{f}\frac{\partial f}{\partial a} = -\frac{\hbar^2}{2\lambda^2}\frac{1}{f}\frac{\partial^2f}{\partial\alpha^2} + e^\alpha  = \omega^2 \ee

\noindent
which is now separable, $\omega^2$ being the separation constant. Now choosing $f(a,\alpha) = A(a)B(\alpha)$ the above equation reads,
\be\label{21} \left\{
  \begin{array}{l}
    \frac{d^2B}{d\alpha^2} + \frac{2\lambda^2}{\hbar^2}(\omega^2 - e^\alpha)B(\alpha) = 0\vspace{.2 in } \\
    \frac{1}{A(a)}\frac{d A(a)}{d a} = \pm\frac{\omega^2}{\hbar\sqrt k}\frac{1}{a^3}
  \end{array}
\right. \ee

\noindent
Solving the first equation of (\ref{21}) we get the following expression,
\be B(\alpha) = C_{29}(-1)^{\frac{i \sqrt 2 \omega\lambda}{\hbar}}I_x(y)\Gamma\left( 1 + x\right)+C_{30}(-1)^{\frac{- i \lambda\sqrt 2 \omega}{\hbar}}I_{-x}(y)\Gamma\left( 1 - x\right)  \ee

\noindent
where $C_{29}$, $C_{30}$ are constants of integration, $ I_\alpha(x) $ is the modified Bessel function of first kind while
\be x = \frac{2\lambda\omega i\sqrt2}{\hbar} \hspace{ .5 in} \mathrm{and}  \hspace{.5 in} y = \frac{2\lambda\sqrt{2e^\alpha}}{\hbar} \ee

\noindent
As in the quantum case here again it is important to note that the sign of $\lambda$ does not affect the solution $B(\alpha)$. Solution of the second equation of (\ref{21}) is,
\be A(a) = A_0 e^{\mp \frac{\omega^2}{2\hbar\sqrt ka^2}} \ee

\noindent
where $ A_0 $ is the integration constant. So the corresponding wavefunction takes the form,
\be \Psi(a,\alpha) = A_0 \mu a^{-\frac{p+1}{2}}\exp{\left[-\frac{1}{\hbar}\left(\frac{M}{2}\sqrt{k}a^2 + \frac{\omega^2}{2\sqrt ka^2}\right) \right]}B(\alpha). \ee

\noindent
The exponential part of the wavefunction is similar to one obtained for massless scalar field ($V_0 = 0$) and is well behaved, nonetheless the factor $B(\alpha)$ kills the Hawking-Page boundary condition for all values of $p$. Thus semiclassical wormhole also does not exist for exponential potential.

\subsubsection{Back-reaction and the wormhole throat}

Although neither quantum nor the semiclassical wormhole exists for the exponential potential under consideration, nevertheless, the semiclassical wavefunction leaves a nice exponential part suitable to find the back-reaction term. Einstein's equation with such back-reaction term here takes the similar form as massless scalar field, viz.,
\be\label{24} \frac{\dot a^2}{a^2}+\frac{1}{a^2} = \frac{2}{M}<\rho_{\phi}> = \frac{2}{M}\left(\frac{\omega^2}{a^6} - \frac{\omega^4}{2Ma^{10}}\right), \ee
which under Wick rotation reads
\be a_{,\tau}^2 = 1-  \frac{2}{M}\left(\frac{\omega^2}{a^4} - \frac{\omega^4}{2Ma^{8}}\right) \ee
\noindent
and it is apparent that $a_{\tau}^2 > 0$ assuring asymptotic flat Euclidean regime. The radius of the throat is given by,
\be a_0 = \left(\frac{\omega^2}{M}\right)^{\frac{1}{4}}. \ee
\noindent
The strong energy condition is satisfied provided,
\be V_0 \le \left(\frac{M\sqrt M}{3 \omega}\right) e^{\frac{\phi}{\lambda}},\ee
which is a reasonable condition for both the signatures of $\lambda$. Thus even in the absence of a quantum or semiclassical wormhole solution the back-reaction under semiclassical approximation leads to a well behaved classical wormhole. The late time cosmic evolution with such wormhole initial condition is determined by the similar set of equations (49) and (50) presented in the case of massless scalar field and so wormhole initial condition does not alter Friedmann solutions.

\section{Concluding remarks}
Euclidean wormholes, which connect two asymptotically flat/de-Sitter space with a throat, are the solutions of Einstein's field equations under Wick rotation. Such macroscopic wormholes are realized for radiation dominated and matter (pressureless dust) dominated era as demonstrated in equations (\ref{fr2}) and (\ref{fm2}), but not for real scalar field. For microscopic wormholes, one has to probe very early universe, which is vacuum dominated or might contain a scalar field. Hawking-Page boundary condition is useful to find such microscopic wormholes. We have explored the possibility of existence of such microscopic wormholes by solving W-D equation corresponding to Einstein-Hilbert action being minimally coupled to a scalar field with varied type of potentials. We have also shown that a scalar can only leave behind a trace which might affect classical cosmological evolution in the form of back-reaction obtained under semiclassical approximation. In the process, $<\rho_{\phi}>$ becomes a function of the scale factor and possibility of obtaining classical Euclidean wormhole solutions for a real scalar field emerge. The findings are the following.\\
1. All type of potentials do not admit Hawking-Page wormhole boundary condition.\\
2. Potentials in the form $V(\phi) = 0, V_0, V_0 \phi^2$ and $V_0\phi^{-1}$ have been found to exhibit both quantum and semiclassical wormhole configurations. Further, classical wormholes under back-reaction exist for all except the inverse potential ($V_0\phi^{-1}$). Additionally, classical cosmological evolution under wormhole initial condition (put up by semiclassical back-reaction) has been found to remain unaltered from Friedmann solution in the radiation and matter dominated era, for massless $V(\phi) = 0$, massive $V(\phi) = V_0 \phi^2$ and exponential $V(\phi) = V_0 e^{-\frac{\phi}{\lambda}}$ scalar fields. The constant potential case leaves a negative cosmological constant under back reaction and so classical cosmological evolution is unrealistic.\\
3. For the potential in the form $V = V_0\phi^{-4}$ quantum wormhole does not exist. Although it admits semiclassical wormhole for $p \le -7$, back-reaction does not give a real throat. Hence, the scalar in this case, does not leave behind a trace for classical wormhole to exist.\\
4. Neither quantum nor semiclassical wormhole exists for exponential potential. Nevertheless, back-reaction term leads to a well behaved classical wormhole solution. Taking into account such wormhole boundary condition, late stage of cosmological evolution has been found to remain unchanged from the Friedmann solutions in the radiation and matter dominated era.\\
5. Potentials in the form $V = V_0 \phi^4$ and $V = V_0\phi^{-3}$ do not admit any of the wormhole configurations.\\
6. Wormhole fixes the curvature parameter $k = +1$ in the early universe.\\
7. Semiclassical wormholes for zero ($V(\phi) = 0$) and constant ($V(\phi) = V_0$) scalar fields are found for arbitrary operator ordering index $p$. Nevertheless, in general, wormholes if exist, requires the operator ordering index $p \le 1$. Quantum wormhole obtained for massless and constant scalar fields require $p = 1$. Thus, wormhole boundary condition also fixes the factor ordering parameter and $p = 1$ may be chosen in general.\\
A renormalized theory of gravity and string effective action under weak field approximation require higher order curvature invariant terms in the gravitational action, which have not been considered here. The wormhole configuration may change considerably if such quantum corrections are incorporated. This we pose in future.

\appendix
\section{A brief account of the role of special functions used in the literature.}

We have obtained solutions in the literature in terms of some special functions, viz., Modified Bessel, Parabolic cylinder, Airy and Confluent hypergeometric functions. Not all the readers are conversant with the properties of these special functions, although these are available in any standard text books on Mathematical methods. To make the present work self-consistent, we brief underlying properties of these special functions.

\subsection{Modified Bessel functions :}
Modified Bessel functions of first kind $I_\alpha$ and second kind $K_\alpha$ are linearly independent solutions to the differential equation
\be\label{B1} x^2 \frac{d^2y}{dx^2} + x \frac{dy}{dx} - (x^2 + \alpha^2)y = 0 \ee
and are defined as
\be I_\alpha(x)=i^{-\alpha}J_\alpha(i x) = \sum_{m=0}^\infty \frac{1}{m!~\Gamma(m+\alpha+1)}\left(\frac{x}{2}\right)^{2m+\alpha}\;\;\;\mathrm{and}\;\;\;
K_\alpha(x)=\frac{\pi}{2}\frac{I_{-\alpha}(x)-I_\alpha(x)}{\sin{(\alpha\pi})}.\ee
\noindent
where, $J_\alpha(i x)$ is the ordinary Bessel functions of imaginary argument. Therefore, unlike ordinary Bessel functions which are oscillating as functions of a real argument, Modified Bessel functions are functions of imaginary argument and so, $I_\alpha(x)$ and $K_\alpha(x)$ are exponentially growing and decaying functions, respectively. $I_\alpha(x)$ vanishes at the origin $x = 0$ for $\alpha > 0$ but remains finite for $\alpha = 0$, and then grows exponentially with $x$. Analogously, $K_\alpha(x)$ diverges at $x = 0$ and exponentially decays as $x$ increases.\\
\noindent
Now, let us turn our attention to equation (36) which under the choice $x = \frac{a^2 M \sqrt k}{2\hbar}$, becomes

\be\label{B2}  x^2 \frac{d^2A}{dx^2} + \left(\frac{p+1}{2}\right) x \frac{dA}{dx} - \left(x^2 + \frac{M \omega^2}{2\hbar^2}\right)A = 0. \ee
For $p=1$, equation (\ref{B2}) takes exactly the same form as equation (\ref{B1}). But for arbitrary $p$, the solution presented in equation (37), is a combination of $I_\alpha(x)$ and gamma function. Due to the diverging behavior of $I_\alpha(x)$, Hawking-Page boundary condition is not satisfied for the solutions of $\Psi$, presented in (37), (72), (87), (92), (129), (131) and (136). But, as the behavior of $K_\alpha(x)$ is just opposite to that of $I_\alpha(x)$, so in equation (105), the term $K_1\left(\frac{\sqrt{8\eta V_0}}{\hbar}\right)$ being multiplied with $\sqrt \eta = a^3\sqrt {\phi}$ is finite at $\eta = 0$, i.e, $a = 0$ and kills IR divergence of equation (72). So, wormhole boundary condition is satisfied for this case. Although similar term appears in (113), but due to the presence of $\sqrt\eta=\frac{\sqrt\phi}{a^3}$ term, unavoidable UV divergence appears at $a=0$.

\subsection{Airy function}\label{A4}
Airy functions $Ai(x)$ and $Bi(x)$, named after the British astronomer George Biddell Airy, are solutions to the Airy differential equation

\be\label{Airy equation} \frac{d^2y}{dx^2}-xy=0,\ee
For real values of $x$, the Airy function can be defined by the improper Riemann integral
\be Ai(x) = \frac{1}{\pi}\int_0^\infty \cos\left(\frac{t^3}{3}+xt\right)dt.\ee
The Airy function of the second kind, denoted $Bi(x)$, is defined as the solution with the same amplitude of oscillation as $Ai(x)$ as $x \rightarrow\infty$ which differs in phase by $\pi/2$, viz.,
\be Bi(x) = \frac{1}{\pi}\int_0^\infty \left[\exp\left(-\frac{t^3}{3}+xt\right)+\sin\left(\frac{t^3}{3}+xt\right)\right]dt.\ee
\noindent
When $x$ is positive, $Ai(x)$ is positive, convex, and decreasing exponentially to zero, while $Bi(x)$ is positive, convex, and increasing exponentially. When $x$ is negative, $Ai(x)$ and $Bi(x)$ oscillate around zero with ever-increasing frequency and slowly ever-decreasing amplitude which is never damped.\\
Now under the choice $x=\frac{2^{\frac{2}{3}}\left(\frac{kM^2}{4\hbar^2} - \frac{a^2MV_0}{2\hbar^2}\right)}{\left(\frac{-M V_0}{\hbar^2}\right)^{\frac{2}{3}}}$, equation (52) becomes exactly same as equation (\ref{Airy equation}), whose solution is given in equation (53). However, $x$ is clearly negative, due to the presence of $a^2$ in the second term and so the function is highly oscillatory.

\subsection{Parabolic cylinder function}\label{A3}

Parabolic cylinder functions [$D_\nu(x)$] and [$D_{-\nu-1}(ix]$) are two independent solutions to the Weber differential equation
\be y''(x)+\left(\nu+\frac{1}{2}-\frac{1}{4}x^2\right)y(x)=0.\ee
The two independent solutions are given by $y=D_\nu(x)$ and $y=D_{-\nu-1}(ix)$, where
\[ D_\nu(x)=2^{\nu/2+1/4}x^{-1/2}W_{\nu/2+1/4,-1/4}\Big(\frac{1}{2}x^2\Big)\]
\be~~~~~~~~=\frac{2^{\nu/2}e^{x^2/4}(-ix)^{1/4}(ix)^{1/4}}{\sqrt x}U\left(-\frac{1}{2}\nu,\frac{1}{2},\frac{1}{2}x^2\right),\ee
where $W_{km}(x)$ is the Whittaker function and $U(a,b,x)$ is a confluent hypergeometric function of the first kind. This function is implemented in Mathematica as $ParabolicCylinderD[\nu, z]$ and it is a regular function. For $\nu$ a nonnegative integer $n$, the solution $D_n$ reduces to
\be D_n(x)=2^{-n/2}e^{x^2/4}H_n\left(-\frac{x}{\sqrt2}\right),\ee
where $H_n(x)$ is a Hermite polynomial. For positive $\nu$, $D_\nu(x)$ is oscillatory but converges for while for negative $\nu$, $D_\nu(x)$ starts from some finite value and converges rapidly.\\
Now for the choice $n=1, m=\frac{3}{2}$, we have form equation (71)

\be B_{,\eta\eta} - \frac{2V_0}{\hbar^2}\eta^2 B - \frac{2\omega^2}{\hbar^2} B=0,\ee
which under transformation $x=\frac{(8V_0)^{\frac{1}{4}}}{\sqrt \hbar}$ becomes

\be B_{,xx} + \left[\left(-\frac{1}{2}-\frac{\omega^2}{\hbar \sqrt{2V_0}}\right)+\frac{1}{2} - \frac{x^2}{4}\right] B =0,\ee
whose solution is

\be\label{P1} B(\eta)= C_{31} {D}_{\left(-\frac{1}{2}-\frac{\omega^2}{\hbar \sqrt{2V_0}}\right)}\left(\frac{(8 V_{0})^\frac{1}{4}~ \eta}{\sqrt \hbar}\right) + C_{32}{D}_{\left(-\frac{1}{2}-\frac{\omega^2}{\hbar \sqrt{2V_0}}\right)}\left(i\frac{(8 V_{0})^\frac{1}{4}~ \eta}{\sqrt \hbar}\right). \ee
But the solution for $A$ part is possible only for $\omega^2=0$, given in equation (72). So under the choice $\omega^2=0$, equation (\ref{P1}) is exactly same as equation (73). Since $\nu=-\frac{1}{2}$ here, the parabolic cylinder function controls the IR divergence
appearing in modified Bessel function in equation (72). Thus wormhole exists for massive scalar field.

\subsection{Confluent Hypergeometric Functions (Kummer's Function)}

Confluent hypergeometric equation
\be xy'' + (b - x)y' - ay = 0,\ee
is obtained from hypergeometric equation by merging two of its singularities. It has a regular singularity at $ x=0 $ and one irregular singularity at $ x=\infty $. The independent solutions of the above equation are called confluent hypergeometric functions of first [$ M(a,b;x)={_1F_1(a,b;x)}$] and second [$U(a,b;x)$] kind. In terms of the Pochhammer symbols these are expressed as,
\be M(a,b;x)={_1F_1(a,b;x)}=\sum_{n=0}^\infty\frac{(a)_n}{(b)_n}\frac{x^n}{ n!}\ee
and
\be U(a,b;x)=\frac{\Gamma(1-b)}{\Gamma(a-b+1)}M(a,b;x)+\frac{\Gamma(b-1)}{\Gamma(a)}M(a-b+1,2-b;x).\ee
$M(a,b;x)$ becomes singular if $b$ becomes negative, otherwise it is a fast increasing function while $U(a,b;x)$ is fast decreasing function.\\ Now, solution (109) of the differential equation (107) for $B(\eta)$ contains product of a linear term in $\eta$ an exponent ($e^{-\omega_0\eta}$) and confluent hypergeometric function. Under the choice $C_{19}=0$, IR divergence of ${_1F_1(a,b;x)}$ disappears, while with $\omega_0 > 0$ the UV divergence of $U(a,b;x)$ is controlled by $\eta = \phi a^6$ term. In the process, wormhole boundary condition is realized for $p \le -1$.

\end{document}